\documentclass[letterpaper,twocolumn,english,journal]{IEEEtran}
\usepackage[T1]{fontenc}
\usepackage[latin9]{inputenc}
\usepackage{color}
\usepackage{babel}
\usepackage{bm}
\usepackage{amsthm}
\usepackage{amsmath}
\usepackage{amssymb}
\usepackage{graphicx}
\usepackage[unicode=true,
 bookmarks=true,bookmarksnumbered=true,bookmarksopen=true,bookmarksopenlevel=1,
 breaklinks=false,pdfborder={0 0 0},backref=false,colorlinks=false]
 {hyperref}
\hypersetup{pdftitle={Your Title},
 pdfauthor={Your Name},
 pdfpagelayout=OneColumn, pdfnewwindow=true, pdfstartview=XYZ, plainpages=false}
\usepackage{breakurl}
\makeatletter

\providecommand{\tabularnewline}{\\}

\theoremstyle{plain}
\newtheorem{thm}{\protect\theoremname}
\theoremstyle{remark}
\newtheorem{rem}[thm]{\protect\remarkname}

\usepackage{cite}
\usepackage{flushend}
\ifCLASSOPTIONcompsoc
\usepackage[caption=false,font=normalsize,labelfont=sf,textfont=sf]{subfig}
\else
\usepackage[caption=false,font=footnotesize]{subfig}
\fi
\usepackage{psfrag,cite}
\usepackage{algorithmic}
\ifCLASSOPTIONcompsoc
\usepackage[caption=false,font=normalsize,labelfont=sf,textfont=sf]{subfig}
\else
\usepackage[caption=false,font=footnotesize]{subfig}
\fi

\@ifundefined{showcaptionsetup}{}{%
 \PassOptionsToPackage{caption=false}{subfig}}
\usepackage{subfig}
\makeatother

\providecommand{\remarkname}{Remark}
\providecommand{\theoremname}{Theorem}

\begin{document}

\title{Spatial Wireless Channel Prediction\\
under Location Uncertainty}

\author{L.~Srikar~Muppirisetty, Tommy Svensson,~\IEEEmembership{Senior~Member,~IEEE},
and Henk~Wymeersch,~\IEEEmembership{Member,~IEEE} %
\thanks{The authors are with the Department of Signals and Systems, Chalmers
University of Technology, Gothenburg, Sweden, e-mail: \{srikar.muppirisetty, henkw, tommy.svensson\}@chalmers.se.

This research was supported, in part, by the European Research Council,
under Grant No. 258418 (COOPNET), and the Swedish Research Council
VR under the project Grant No. 621-2009-4555 (Dynamic Multipoint Wireless
Transmission).%
}}
\maketitle
\begin{abstract}
Spatial wireless channel prediction is important for future wireless
networks, and in particular for proactive resource allocation at different
layers of the protocol stack. Various sources of uncertainty must
be accounted for during modeling and to provide robust predictions.
We investigate two channel prediction frameworks, classical Gaussian
processes (cGP) and uncertain Gaussian processes (uGP), and analyze
the impact of location uncertainty during learning/training and prediction/testing,
for scenarios where measurements uncertainty are dominated by large-scale
fading. We observe that cGP generally fails both in terms of learning
the channel parameters and in predicting the channel in the presence
of location uncertainties.\textcolor{blue}{{} }In contrast, uGP explicitly
considers the location uncertainty. Using simulated data, we show
that uGP is able to learn and predict the wireless channel.\end{abstract}

\begin{IEEEkeywords}
Gaussian processes, uncertain inputs, location uncertainty, spatial
predictability of wireless channels.
\end{IEEEkeywords}

\section{Introduction}

\IEEEPARstart{L}{ocation}-based resource allocation schemes are
expected to become an essential element of emerging 5G networks, as
5G devices will have the capability to accurately self-localize and
predict relevant channel quality metrics (CQM) \cite{sand2009position,location2014rocco,abou2013optimal}
based on crowd-sourced databases. The geo-tagged CQM (including, e.g.,
received signal strength, delay spread, and interference levels) from
users enables the construction of a dynamic database, which in turn
allows the prediction of CQM at arbitrary locations and future times.
Current standards are already moving in this direction through the
so-called minimization of drive test (MDT) feature in 3GPPP Release
10 \cite{johansson2012Minimization}. In MDT, users collect radio
measurements and associated location information in order to assess
network performance. In terms of applications, prediction of spatial
wireless channels (e.g., through radio environment maps) and its utilization
in resource allocation can reduce overheads and delays due to the
ability to predict channel quality beyond traditional time scales
\cite{location2014rocco}. Exploitation of location-aware CQM is relevant
for interference management in two-tier cellular networks \cite{zalonis2012femto},
coverage hole detection and prediction \cite{galindo2013harvesting},
cooperative spectrum sensing in cognitive radios \cite{nevat2012location},
anticipatory networks for predictive resource allocation \cite{abou2013optimal},
and proactive caching \cite{TadrousProactive2013}.

In order to predict location-dependent radio propagation channels,
we rely on mathematical models, in which the physical environment,
including the locations of transmitter and receiver, play an important
role. The received signal power in a wireless channel is mainly affected
by three major dynamics, which occur at different length scales: path-loss,
shadowing, and small-scale fading\cite{goldsmith2005wireless}. Small-scale
fading decorrelates within tens of centimeters (depending on the carrier
frequency), making it infeasible to predict based on location information.
On the other hand, shadowing is correlated up to tens of meters, depending
on the propagation environment (e.g., 50--100 m for outdoor \cite{goldsmith2005wireless}
and 1--2 m for indoor environments\cite{jalden2207correlation}).
Finally, path-loss, which captures the deterministic decay of power
with distance, is a deterministic function of the distance to the
transmitter. In rich scattering environments, the measurements average
small-scale fading either in frequency or space provided sufficient
bandwidth or number of antennas\cite{jalden2207correlation}.\emph{
}Thus, provided that measurements are dominated by large-scale fading,
location-dependent models for path-loss and shadowing can be developed
based on the physical properties of the wireless channel. With the
help of spatial regression tools, these large-scale channel components
can be predicted at other locations and used for resource allocation
\cite{sand2009position}. However, since localization is subject to
various error sources (e.g., the global positioning system (GPS) gives
an accuracy of around 10 m \cite{grewal2001global} in outdoor scenarios,
while ultra-wide band (UWB) systems can give sub-meter accuracy),
there is a fundamental need to account for location uncertainties
when developing spatial regression tools.

Spatial regression tools generally comprise a training/learning phase,
in which the underlying channel parameters are estimated based on
the available training database, and a testing/prediction phase, in
which predictions are made at test locations, given learned parameters
and the training database. Among such tools, Gaussian processes (GP)
is a powerful and commonly used regression framework, since it is
generally considered to be the most flexible and provides prediction
uncertainty information \cite{rasmussen2006gaussian}. Two important
limitations of GP are its computational complexity \cite{csato2002sparse,quinonero2005unifying,snelson2006sparse,sarkka2013spatio}
and its sensitivity to uncertain inputs \cite{Dallaire2011,girard2004approximate,girard2003learning,quinonero2005unifying,jadaliha2013gaussian,mchutchon2011gaussian}.
To alleviate the computational complexity, various sparse GP techniques
have been proposed in \cite{quinonero2005unifying,csato2002sparse,snelson2006sparse},
while online and distributed GP were treated in \cite{Ranganathan2011Online,kim2011cooperative,sarkka2013spatio}
and \cite{Dongbing2012Spatial,deisenroth2015distributed,ChoiDistributed2013},
respectively. The impact of input uncertainty was studied in \cite{Dallaire2011,girard2004approximate},
which showed that GP was adversely affected, both in training and
testing, by input uncertainties. The input uncertainty in our case
corresponds to location uncertainty.

No framework has yet been developed to mathematically characterize
and understand the spatial predictability of wireless channels with
location uncertainty. In this paper, we build on and adapt the framework
from \cite{Dallaire2011,girard2004approximate} to CQM prediction
in wireless networks. Our main contributions are as follows:
\begin{itemize}
\item We show that not considering location uncertainty leads to poor learning
of the channel parameters and poor prediction of CQM values at other
locations, especially when location uncertainties are heterogeneous;
\item We relate and unify existing GP methods that account for uncertainty
during both learning and prediction, by operating directly on an input
set of distributions, rather than an input set of locations;
\item We describe and delimit proper choices for mean functions and covariance
functions in this unified framework, so as to incorporate location
uncertainty in both learning and prediction; and
\item We demonstrate the use of the proposed framework for simulated data
and apply it to a spatial resource allocation application.
\end{itemize}
The remainder of the paper is structured as follows. Section \ref{sec:System-Model}
presents the channel model and details the problem description for
location-dependent channel prediction with location uncertainty. In
Section \ref{sec:cGP}, we review channel learning and prediction
in the classical GP (cGP) setup with no localization errors. Section
\ref{sec:uGP} details learning and prediction procedures using the
proposed GP framework that accounts for uncertainty on training and
test locations, termed uncertain GP (uGP). Finally, numerical results
are given in Section \ref{sec:Numerical-Results} in addition to a
resource allocation example, followed by our conclusions in Section
\ref{sec:Conclusion}.

\subsubsection*{Notation}

Vectors and matrices are written in bold (e.g., a vector $\mathbf{k}$
and a matrix\textbf{ $\mathbf{K}$}); $\mathbf{K}{}^{\mathrm{T}}$
denotes transpose of \textbf{$\mathbf{K}$}; $\vert\mathbf{K}\vert$
denotes determinant of \textbf{$\mathbf{K}$}; $[\mathbf{K}]_{ij}$
denotes entry $(i,j)$ of $\mathbf{K}$;\textbf{ $\mathbf{I}$ }denotes
identity matrix of appropriate size; $\mathbf{1}$ and $\mathbf{0}$
are vectors of ones and zeros, respectively, of appropriate size;
$\Vert.\Vert$ denotes $L_{2}$-norm unless otherwise stated; $\mathbb{E}[.]$
denotes the expectation operator; $\textrm{Cov}[.]$ denotes covariance
operator (i.e., $\textrm{Cov}[\mathbf{y}_{1},\mathbf{y}_{2}]=\mathbb{E}[\mathbf{y}_{1}\mathbf{y}_{2}^{\mathrm{T}}]-\mathbb{E}[\mathbf{y}_{1}]\,\mathbb{E}[\mathbf{y}_{2}]^{\mathrm{T}}$);
$\mathcal{N}(\mathbf{x};\mathbf{m},\bm{\Sigma})$ denotes a Gaussian
distribution evaluated in $\mathbf{x}$ with mean vector $\mathbf{m}$
and covariance matrix $\bm{\Sigma}$ and $\mathbf{x}\sim\mathcal{N}(\mathbf{m},\bm{\Sigma})$
denotes that $\mathbf{x}$ is drawn from a Gaussian distribution with
mean vector $\mathbf{m}$ and covariance matrix $\bm{\Sigma}$. Important
symbols used in the paper are: $\mathbf{x}_{i}\in\mathbb{R}^{2}$
is an exact, true location; $\mathbf{u}_{i}\in\mathbb{R}^{D}$, $D>2$
is a vector that describes (e.g., in the form of moments) the location
distribution $p(\mathbf{x}_{i})$. For example in the case of Gaussian
distributed localization error, $p(\mathbf{x})=\mathcal{N}(\mathbf{x};\mathbf{z},\bm{\Sigma})$,
then a possible choice is $\mathbf{u}=[\mathbf{z}^{\mathrm{T}},\textrm{vec}[\bm{\Sigma}]]^{\mathrm{T}}$,
where $\mathrm{vec}[\mathbf{\Sigma}]$ stacks all the elements of
$\mathbf{\Sigma}$ in a vector. Finally, $\mathbf{z}_{i}=\phi(\mathbf{u}_{i})\in\mathbb{R}^{2}$
is a location estimate extracted from $\mathbf{u}_{i}$ through a
function $\phi(\cdot)$ (e.g., the mean or mode).

\section{Related Work}

First, we give an overview of the literature on GP with uncertain
inputs. One way to deal with the input noise is through linearizing
the output around the mean of the input \cite{mchutchon2011gaussian,girard2003learning}.
In \cite{mchutchon2011gaussian}, the input noise was viewed as extra
output noise by linearization at each point and this is proportional
to the squared gradient of the GP posterior mean. However, the proposed
method works under the condition of constant-variance input noise.
In \cite{girard2003learning}, a Delta method was used for linearization
under the assumption of Gaussian distributed inputs and proposed a
corrected covariance function that accounts for the input noise variance.
For Gaussian distributed test inputs and known training inputs, the
exact and approximate moments of the GP posterior was examined for
various forms of covariance functions \cite{girard2004approximate}.
Training on Gaussian distributed input points by calculating the expected
covariance matrix was studied in \cite{Dallaire2011,girard2004approximate}.
Two approximations were evaluated in \cite{quinonero2004learning},
first a joint maximization of joint posterior on uncertain inputs
and hyperparameters (leading to over-fitting), and second using a
stochastic expectation--maximization algorithm (at a high computational
cost).

We now review previous works on GP for channel prediction, which include
spatial correlation of shadowing in cellular \cite{fink2011communication}
and ad-hoc networks \cite{agrawal2009correlated}, as well as tracking
of transmit powers of primary users in a cognitive network \cite{kim2011cooperative}.
In \cite{fink2011communication}, GP was shown to model spatially
correlated shadowing to predict shadowing and path-loss at any arbitrary
location. A multi-hop network scenario was considered \cite{agrawal2009correlated},
and shadowing was modeled using a spatial loss field, integrated along
a line between transmitter and receiver. In \cite{kim2011cooperative},
a cognitive network setting was evaluated, in which the transmit powers
of the primary users were tracked with cooperation among the secondary
users. For this purpose a distributed radio channel tracking framework
using Kriged Kalman filter was developed with location information.
A study on the impact of underlying channel parameters on the spatial
channel prediction variance using GP was presented in \cite{malmirchegini2012spatial}.
A common assumption in \cite{fink2011communication,agrawal2009correlated,kim2011cooperative,malmirchegini2012spatial}
was the presence of perfect location information. This assumption
was partially removed in \cite{yanimpact}, which extends \cite{malmirchegini2012spatial}
to include the effect of localization errors on spatial channel prediction.
It was found that channel prediction performance was degraded when
location errors were present, in particular when either the shadowing
standard deviation or the shadowing correlation were large. However,
\cite{yanimpact} did not tackle combined learning and prediction
under location uncertainty. The only work that explicitly accounts
for location uncertainty was \cite{jadaliha2013gaussian}, in which
the Laplace approximation was used to obtain a closed-form analytical
solution for the posterior predictive distribution. However, \cite{jadaliha2013gaussian}
did not consider learning of parameters in presence of location uncertainty.

\section{System Model \label{sec:System-Model}}

\subsection{Channel Model\label{sub:Channel-Model}}

Consider a geographical region $\mathcal{A}\subseteq\mathbb{R}^{2}$,
where a source node is located at the origin and transmits a signal
with power $P_{\mathrm{TX}}$ to a receiver located at $\mathbf{x}_{i}\in\mathcal{A}$
through a wireless propagation channel. The received radio signal
is affected mainly by distance-dependent path-loss, shadowing due
to obstacles in the propagation medium, and small-scale fading due
to multipath effects. The received power $P_{\mathrm{RX}}(\mathbf{x}_{i})$
can be expressed as \cite[Chap. 2]{Stuber2001principle}
\begin{equation}
P_{\mathrm{RX}}(\mathbf{x}_{i})=P_{\mathrm{TX}}\, g_{0}\,\Vert\mathbf{x}_{i}\Vert^{-\eta}\,\psi(\mathbf{x}_{i})\,|h(\mathbf{x}_{i})|^{2},
\end{equation}
where $g_{0}$ is a constant that captures antenna and other propagation
gains, $\eta$ is the path-loss exponent, $\psi(\mathbf{x}_{i})$
is the location-dependent shadowing and $h(\mathbf{x}_{i})$ is the
small-scale fading. We assume measurements average%
\footnote{If measurements cannot average over small-scale fading, the proposed
framework from this paper cannot be applied.\textcolor{blue}{{} }%
} small-scale fading, either in time (measurements taken over a time
window), frequency (measurements represent average power over a large
frequency band), or space (measurements taken over multiple antennas)
\cite{Goldsmith1994Error,jalden2207correlation}. Therefore, the resulting
received signal power from the source node to a receiver node $i$
can be expressed in dB scale as
\begin{equation}
P_{\mathrm{RX}}(\mathbf{x}_{i})\mbox{[dBm]}=L_{0}-10\,\eta\,\log_{10}(\Vert\mathbf{x}_{i}\Vert)+\Psi(\mathbf{x}_{i}),\label{eq:loc_dep_rec_pow}
\end{equation}
where $L_{0}=P_{\mathrm{TX}}\mbox{[dBm]}+G_{0}$ with $G_{0}=10\,\log_{10}(g_{0})$
and $\Psi(\mathbf{x}_{i})=10\,\log_{10}(\psi(\mathbf{x}_{i}))$. A
common choice for modeling shadowing in wireless systems is through
a log-normal distribution, i.e., $\Psi(\mathbf{x}_{i})\sim\mathcal{N}(0,\sigma_{\Psi}^{2})$,
where $\sigma_{\Psi}^{2}$ is the shadowing variance. Shadowing $\Psi(\mathbf{x}_{i})$
is spatially correlated, with well-established correlation models
\cite{szyszkowicz2010feasibility}, among which the Gudmundson model
is widely used \cite{gudmundson1991correlation}. Let $y_{i}$ be
the scalar%
\footnote{Vector measurements are also possible (e.g., from multiple base stations),
but not considered here for the sake of clarity.%
} observation of the received power at node $i$, which is written
as $y_{i}=P_{\mathrm{RX}}(\mathbf{x}_{i})+n_{i},$ where $n_{i}$
is a zero mean additive white Gaussian noise with variance $\sigma_{n}^{2}$.
For the sake of notational simplicity, we do not consider a three-dimensional
layout, the impact of non-uniform antenna gain patterns, or distance-dependent
path-loss exponents.

\subsection{Location Error Model}

In practice, nodes may not have access to their true location $\mathbf{x}_{i}$,
but only to a distribution $p(\mathbf{x}_{i})$\footnote{$p(\mathbf{x}_{i})$ is used for $p(\mathbf{x}=\mathbf{x}_{i})$ for
notational simplicity.}. The distribution $p(\mathbf{x}_{i})$ is obtained
from the positioning algorithm in the devices, and depends on the
specific positioning technology (e.g., for GPS the distribution $p(\mathbf{x}_{i})$
can be modeled as a Gaussian). We will assume that all distributions
$p(\mathbf{x}_{i})$ come from a given family of distributions (e.g.,
all bivariate Gaussian distributions). These distributions can be
described by a finite set of parameters, $\mathbf{u}_{i}\in\mathbb{R}^{D}$,
$D>2$, e.g., a mean and a covariance matrix for Gaussian distributions.
The set of descriptions of all distributions from the given family
is denoted by $\mathcal{U}\subset\mathbb{R}^{D}$. Within this set,
the set of all delta Dirac distributions over locations is denoted
by $\mathcal{X}\subset\mathcal{U}$. Note that $\mathcal{X}$ is equivalent
to the set $\mathcal{A}$ of possible locations. Finally, we introduce
a function $\phi:\mathcal{U}\to\mathcal{A}$ that extracts a position
estimate from the distribution (in our case chosen as the mean), and
denote $\mathbf{z}_{i}=\phi(\mathbf{u}_{i})\in\mathcal{A}$. We will
generally make no distinction between a distribution $p(\mathbf{x}_{i})$
and its representation $\mathbf{u}_{i}$.

\subsection{Problem Statement}

\begin{figure}[t]
\psfrag{cGP}{\hspace{-5mm} \footnotesize{classical GP}}
\psfrag{uGP}{\hspace{-5mm}  \footnotesize{uncertain GP}}
\psfrag{z}{\footnotesize{$\{ \mathbf{Z},\mathbf{y} \}, \mathbf{z}^*$}}
\psfrag{u}{\footnotesize{$\{ \mathbf{U},\mathbf{y} \}, \mathbf{u}^*$}}

\includegraphics[width=1\columnwidth]{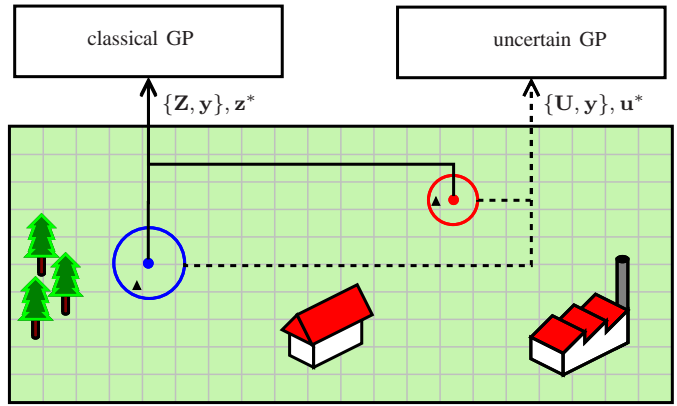}\protect\caption{\textcolor{blue}{\label{fig:cGP_uGP_highlevel}}High-level comparison
between cGP and uGP. The inputs to cGP during learning are observations
$\mathbf{Y}$ and estimates $\mathbf{Z}$ of the (unobserved) actual
locations $\mathbf{X}$ where those observations have been taken.
$\mathbf{Z}$ is obtained through a positioning system. The true locations
$\mathbf{X}$ are marked with a triangle and are generally different
from the estimated locations $\mathbf{Z}$, marked with a blue and
red dot. During prediction, cGP predicts received power at an estimated
test location, $\mathbf{z}^{*}$. In contrast, uGP considers the distribution
of the locations $\mathbf{X}$, described by $\mathbf{U}$ (and depicted
by the red and blue circle), during learning. During prediction, uGP
utilizes the distribution $\mathbf{u}^{*}$ of the test location.
Note that the amount of uncertainty (radius of the circle) can change. }
\end{figure}

We assume a central coordinator, which collects a set of received
power measurements $\mathbf{y}=[y_{1},\ldots,y_{N}]^{T}$ with respect
to a common source from $N$ nodes, along with their corresponding
location distributions $\mathbf{U}=[\mathbf{u}_{1}^{\mathrm{T}},\mathbf{u}_{2}^{\mathrm{T}},\ldots,\mathbf{u}_{N}^{\mathrm{T}}]^{\mathrm{T}}$.
Our goals are to perform
\begin{enumerate}
\item \emph{Learning}: construct a spatial model (through estimating model
parameters $\boldsymbol{\theta}$, to be defined later) of the received
power based on the measurements;
\item \emph{Prediction}: determine the predictive distribution $p(P_{\mathrm{RX}}(\mathbf{x}_{*})|\mathbf{y},\mathbf{U},\hat{\boldsymbol{\theta}},\mathbf{x}_{*})$
of the power in test locations $\mathbf{x}_{*}$ and the distribution
of the expected%
\footnote{Here, $P_{\mathrm{RX}}(\mathbf{u}_{*})$ should be interpreted as
the expected received power, $p(P_{\mathrm{RX}}(\mathbf{u}_{*})|\mathbf{y},\mathbf{U},\hat{\boldsymbol{\boldsymbol{\theta}}},\mathbf{u}_{*})=\int p(P_{\mathrm{RX}}(\mathbf{x}_{*})|\mathbf{y},\mathbf{U},\hat{\boldsymbol{\boldsymbol{\theta}}},\mathbf{x}_{*})p(\mathbf{x}_{*})\mathrm{d}\mathbf{x}_{*}$,
where $p(\mathbf{x}_{*})$ is described by $\mathbf{u}_{*}$%
} received power, $p(P_{\mathrm{RX}}(\mathbf{u}_{*})|\mathbf{y},\mathbf{U},\hat{\boldsymbol{\theta}},\mathbf{u}_{*})$,
for test location distributions $\mathbf{u}_{*}$.
\end{enumerate}
We will consider two methods for learning and prediction: classical
GP (Section \ref{sec:cGP}), which ignores location uncertainty and
only considers $\mathbf{z}_{i}=\phi(\mathbf{u}_{i})$, and uncertain
GP (Section \ref{sec:uGP}), which is a method that explicitly accounts
for location uncertainty. We introduce $\mathbf{X}=[\mathbf{x}_{1}^{\mathrm{T}},\mathbf{x}_{2}^{\mathrm{T}},\ldots,\mathbf{x}_{N}^{\mathrm{T}}]^{\mathrm{T}}$
and $\mathbf{Z}=[\mathbf{z}_{1}^{\mathrm{T}},\mathbf{z}_{2}^{\mathrm{T}},\ldots,\mathbf{z}_{N}^{\mathrm{T}}]^{\mathrm{T}}$
as the collection of true and estimated locations respectively. A
high level comparison of cGP and uGP is shown in Fig.~\ref{fig:cGP_uGP_highlevel},
where cGP operates on $\mathbf{Z}$ and $\mathbf{Y}$, while uGP operates
on $\mathbf{U}$ and $\mathbf{Y}$.

\section{Channel Prediction with Classical GP\label{sec:cGP}}

We first present cGP under the assumption that all locations during
learning and prediction are known exactly, based on \cite{rasmussen2006gaussian,mostofi2010estimation}.
Later in this section, we will discuss the impact of location uncertainties
on cGP in learning/training and prediction/testing.

\subsection{cGP without Location Uncertainty \label{sub:Classical-GP-Description}}

We designate $\mathbf{x}_{i}\in\mathcal{A}$ as the \emph{input} variable,
and $P_{\mathrm{RX}}(\mathbf{x}_{i})$ as the \emph{output }variable.
We model $P_{\mathrm{RX}}(\mathbf{x}_{i})$ as a GP with mean function
$\mu(\mathbf{x}_{i}):\,\mathcal{A}\rightarrow\mathbb{R}$ and a positive
semidefinite covariance function $C(\mathbf{x}_{i},\mathbf{x}_{j}):\,\mathcal{A}\times\mathcal{A}\rightarrow\mathbb{R}^{+}$,
and we write
\begin{equation}
{\normalcolor {\color{red}{\normalcolor P_{\mathrm{RX}}(\mathbf{x}_{i})\sim\mathcal{GP}(\mu(\mathbf{x}_{i}),C(\mathbf{x}_{i},\mathbf{x}_{j})),}}}
\end{equation}
where $\mathcal{GP}$ stands for a Gaussian process. The mean function%
\footnote{Other ways of including the mean function in the model are possible,
such as to include it in the covariance structure, and transform the
prior model to a zero-mean GP prior \cite{rasmussen2006gaussian}. %
} is defined as $\mu(\mathbf{x}_{i})=\mathbb{E}_{\Psi(\mathbf{x}_{i})}[P_{\mathrm{RX}}(\mathbf{x}_{i})]=L_{0}-10\,\eta\;\log_{10}(\Vert\mathbf{x}_{i}\Vert)$,
due to (\ref{eq:loc_dep_rec_pow}). The covariance function is defined
as $C(\mathbf{x}_{i},\mathbf{x}_{j})=\mathrm{Cov}[P_{\mathrm{RX}}(\mathbf{x}_{i}),P_{\mathrm{RX}}(\mathbf{x}_{j})]$.
We will consider a class of covariance functions of the form:
\begin{equation}
C(\mathbf{x}_{i},\mathbf{x}_{j})=\sigma_{\Psi}^{2}\exp\left(-\frac{\Vert\mathbf{x}_{i}-\mathbf{x}_{j}\Vert{}^{p}}{d_{c}^{\, p}}\right)+\delta_{ij}\,\sigma_{\mathrm{proc}}^{2},\label{eq:generic_cov_function}
\end{equation}
where $\delta_{ij}=1$ for $i=j$ and zero otherwise, $p\ge1$, $d_{c}$
is the correlation distance of the shadowing, and $\sigma_{\mathrm{proc}}$
captures any noise variance term that is not due to measurement noise
(more on this later). Setting $p=1$ in (\ref{eq:generic_cov_function}),
gives the exponential covariance function that is commonly used to
describe the covariance properties of shadowing \cite{gudmundson1991correlation},
and $p=2$, gives the squared exponential covariance function that
will turn out to be useful in Section \ref{sub:Uncertain-GP}. Note
that the mean and covariance depend on
\begin{equation}
\boldsymbol{\theta}=[\sigma_{n},\sigma_{\mathrm{proc}},d_{c},L_{0},\eta,\sigma_{\Psi}],
\end{equation}
which may not be known a priori.

\subsubsection{Learning}

The objective during learning is to infer the model parameters $\boldsymbol{\theta}$
from observations $\mathbf{y}$ of the received power at $N$ \emph{known}
locations $\mathbf{X}$. The resulting training database is thus $\{\mathbf{X},\mathbf{y}\}$.
Due to the GP model, the joint distribution of the $N$ training observations
exhibits a Gaussian distribution
\begin{align}
p(\mathbf{y}\vert\mathbf{X},\boldsymbol{\theta}) & \mathbf{=}\mathcal{N}(\mathbf{y};\mathbf{\mathbb{\boldsymbol{\mu}}}(\mathbf{X}),\mathbf{K}),\label{eq:GPtraining}
\end{align}
where $\mathbf{\boldsymbol{\mathbf{\mu}}}(\mathbf{X})=[\mu(\mathbf{x}_{1}),\mu(\mathbf{x}_{2}),\ldots,\mu(\mathbf{x}_{N})]^{\mathrm{T}}$
is the mean vector and $\mathbf{K}$ is the covariance matrix of the
measured received powers, with entries $[\mathbf{K}]_{ij}=C(\mathbf{x}_{i},\mathbf{x}_{j})+\sigma_{n}^{2}\,\delta_{ij}$.
The model parameters can be learned through maximum likelihood estimation,
given the training database $\{\mathbf{X},\mathbf{y}\}$, by minimizing
the negative log-likelihood function with respect to $\boldsymbol{\theta}$:
\begin{align}
\hat{\boldsymbol{\theta}} & =\arg\underset{\boldsymbol{\theta}}{\min}\{-\log(p(\mathbf{y}\vert\mathbf{X},\boldsymbol{\theta}))\}.\label{eq:learning_cGP}
\end{align}
The negative log-likelihood function is usually not convex and may
contain multiple local optima. Additional details on the learning
process are provided later. Once $\hat{\boldsymbol{\theta}}$ is determined
from $\{\mathbf{X},\mathbf{y}\}$, the training process is complete.

\subsubsection{Prediction}

After learning, we can determine the predictive distribution of $P_{\mathrm{RX}}(\mathbf{x}_{*})$
at a new and arbitrary test location $\mathbf{x}_{*}$, given the
training database $\{\mathbf{X},\mathbf{y}\}$ and $\hat{\boldsymbol{\theta}}$.
We first form the joint distribution
\begin{equation}
\left[\begin{array}{c}
\mathbf{y}\\
P_{\mathrm{RX}}(\mathbf{x}_{*})
\end{array}\right]\sim\mathcal{N}\left(\left[\begin{array}{c}
\boldsymbol{\mu}(\mathbf{X})\\
\mu(\mathbf{x}_{*})
\end{array}\right],\left[\begin{array}{cc}
\mathbf{K} & \mathbf{k}_{*}\\
\mathbf{k}_{*}^{\mathrm{T}} & k_{**}
\end{array}\right]\right),\label{eq:gaussian_joint distribution}
\end{equation}
where $\mathbf{k}_{*}$ is the $N\times1$ vector of cross-covariances
$C(\mathbf{x_{*}},\mathbf{x}_{i})$ between the received power at
$\mathbf{x}_{*}$ and at the training locations $\mathbf{x}_{i}$,
and $k_{**}=C(\mathbf{x}_{*},\mathbf{x}_{*})$ is the prior variance
(i.e., the variance in the absence of measurements), given by $C(\mathbf{x}_{*},\mathbf{x}_{*})$.
Conditioning on the observations\textbf{ $\mathbf{y}$}, we obtain
the Gaussian posterior distribution $p(P_{\mathrm{RX}}(\mathbf{x}_{*})\vert\mathbf{X},\mathbf{y},\hat{\boldsymbol{\theta}},\mathbf{x}_{*})$
for the test location $\mathbf{x}_{*}$. The mean ($\bar{P}_{\mathrm{RX}}(\mathbf{x}_{*})$)
and variance ($V_{\mathrm{RX}}(\mathbf{x}_{*})$) of this distribution
turn out to be \cite{rasmussen2006gaussian}
\begin{align}
\bar{P}_{\mathrm{RX}}(\mathbf{x}_{*})= & \mu(\mathbf{x}_{*})+\mathbf{k}_{*}^{\mathrm{T}}\,\mathbf{K}^{-1}\,(\mathbf{y-\boldsymbol{\mu}(\mathbf{X})})\label{eq:mu_pred_classicGP}\\
= & \mu(\mathbf{x}_{*})+\sum_{i,j=1}^{N}[\mathbf{K}{}^{-1}]_{ij}\,(y_{j}-\mu(\mathbf{x}_{j}))\, C(\mathbf{x}_{*},\mathbf{x}_{i})\nonumber \\
= & \mu(\mathbf{x}_{*})+\sum_{i=1}^{N}\beta_{i}\, C(\mathbf{x}_{*},\mathbf{x}_{i}).\nonumber \\
V_{\mathrm{RX}}(\mathbf{x}_{*})= & k_{**}-\mathbf{k}_{*}^{\mathrm{T}}\,\mathbf{K}^{-1}\,\mathbf{k}_{*}\label{eq:var_pred_classicGP}\\
= & k_{**}-\sum_{i,j=1}^{N}[\mathbf{K}{}^{-1}]_{ij}\, C(\mathbf{x}_{*},\mathbf{x}_{i})\, C(\mathbf{x}_{*},\mathbf{x}_{j}),\nonumber
\end{align}
where $\beta_{i}=\sum_{j=1}^{N}[\mathbf{K}{}^{-1}]_{ij}(y_{j}-\mu(\mathbf{x}_{j}))$.
In (\ref{eq:mu_pred_classicGP}), $\mu(\mathbf{x}_{*})$ corresponds
to the deterministic path-loss component at $\mathbf{x}_{*}$, which
is corrected by a term involving the database and the correlation
between the measurements at the training locations and the test location.
In (\ref{eq:var_pred_classicGP}), we see that the prior variance
$k_{**}$ is reduced by a term that accounts for the correlation of
nearby measurements.
\begin{figure}[t]
\includegraphics[width=1\columnwidth]{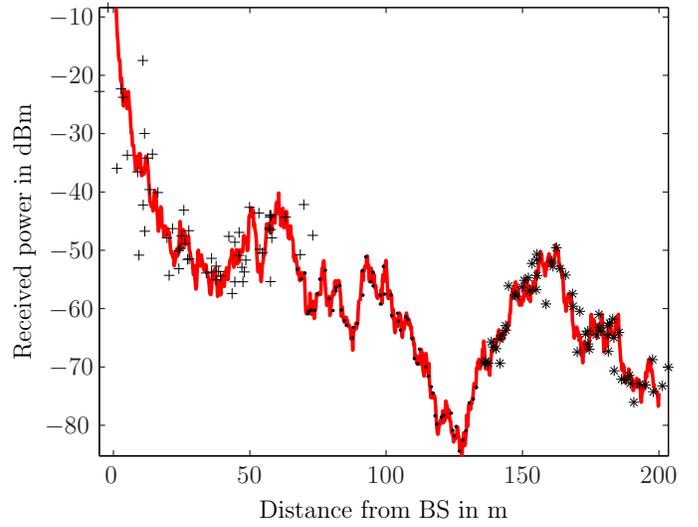}

\protect\caption{\label{fig:Training-locations-with}Impact of location uncertainty
for a one-dimensional example: the red curve depicts the received
signal power $P_{\mathrm{RX}}(\mathbf{x})$ as a function of $\mathbf{x}$
(or equivalently, the distance to the base station), while the markers
show $ $$P_{\mathrm{RX}}(\mathbf{x}_{i})$ as a function of $\mathbf{z}_{i}=\phi(\mathbf{u}_{i})$.
Training measurements are grouped into three regions: (+) corresponds
to high uncertainty, ($\cdot$) corresponds to low uncertainty, and
({*}) corresponds to medium uncertainty, respectively. The location
uncertainty results in output noise.}
\end{figure}

\subsection{cGP with Location Uncertainty}

Now let us consider the case when the nodes do not have access to
their true location $\mathbf{x}_{i}$, but only to a distribution
$p(\mathbf{x}_{i})$, which is described by $\mathbf{u}_{i}\in\mathcal{U}$.
Fig.~\ref{fig:Training-locations-with} illustrates the impact of
location uncertainties assuming Gaussian location errors for a one-dimensional
example. The figure shows (in red) the true received power $P_{\mathrm{RX}}(\mathbf{x})$
as a function of $\mathbf{x}$ as well as the measured power $P_{\mathrm{RX}}(\mathbf{x}_{i})$
as a function of $\mathbf{z}_{i}=\phi(\mathbf{u}_{i})$ for a discrete
number of values of $\mathbf{u}$, shown as markers. To clearly illustrate
the impact of different amounts on uncertainty on the position, we
have artificially created three regions: high location uncertainty
close to the transmitter, medium location uncertainty far away, and
low location uncertainty for intermediate distances. When there is
no location uncertainty (70 m until 140 m from the transmitter), $\mathbf{z}_{i}\approx\mathbf{x}_{i}$,
so $P_{\mathrm{RX}}(\mathbf{z}_{i})\approx P_{\mathrm{RX}}(\mathbf{x}_{i})$,
and hence the black dots coincide with the red curve. For medium and
high uncertainty, $\mathbf{z}_{i}$ can differ significantly from
$\mathbf{x}_{i}$, so the data point with coordinates $[\mathbf{z}_{i},P_{\mathrm{RX}}(\mathbf{x}_{i})]$
can lie far away from the red curve, especially for high location
uncertainty (distances below 70 m). From Fig.~\ref{fig:Training-locations-with}
it is clear that the input uncertainty manifests itself as output
noise, with a variance that grows with increasing location uncertainty\footnote{In fact, the output noise induced by location uncertainty will also
depend on the slope of $P_{\mathrm{RX}}(\mathbf{x}_{i})$ around $\mathbf{x}_{i}$,
since a locally flat function will lead to less output noise than
a steep function, under the same location uncertainty.}. This output noise must be accounted for in the model during learning
and prediction. When these uncertainties are ignored, both learning
and prediction will be of poor quality, as described below.

\subsubsection{Learning from uncertain training locations\label{sub:CGP_UL_CT}}

In this case, the training database\textbf{ $\{\mathbf{Z},\mathbf{y}\}$}
comprises  locations $\mathbf{z}_{i}=\phi(\mathbf{u}_{i})$ and power
measurements $y_{i}=P_{\mathrm{RX}}(\mathbf{x}_{i})+n_{i}$ at the
true (but unknown) locations $\mathbf{x}_{i}$. The measurements will
be of the form shown in Fig.~\ref{fig:Training-locations-with}.
The estimated model parameters $\hat{\boldsymbol{\theta}}$ can take
two forms: (i) assign very short correlation distances $\hat{d}_{c}$,
large $\hat{\sigma}_{\Psi}$, and small $\hat{\sigma}_{\mathrm{proc}}$,
as some seemingly nearby events will appear uncorrelated: or (ii)
assign larger correlation distances $\hat{d}_{c}$, smaller $\hat{\sigma}_{\Psi}$,
and explain the measurements by assigning a higher value to $\hat{\sigma}_{\mathrm{proc}}$
\cite{mchutchon2011gaussian}. In the first case, correlations between
measurement cannot be exploited, so that during prediction, the posterior
mean will be close to the prior mean and the posterior variance will
be close to the prior variance. In the second case, predictions will
be better, as correlations can be exploited to reduce the posterior
variance. However, the model must explain different levels of input
uncertainty with a single covariance function, which can make no distinctions
between locations with low, medium, or high uncertainty. This will
lead to poor performance when location error statistics differ from
node to node.

\subsubsection{Prediction at an uncertain test location\label{sub:CGP_CL_UT}}

In the case where training locations are exactly known (i.e., $\mathbf{z}_{i}=\mathbf{x}_{i}$,
$\forall i$), we may want to predict the power at an uncertain test
location $\mathbf{u}_{*}$, made available to cGP in the form $\mathbf{z}_{*}=\phi(\mathbf{u}_{*})$,
while the true test location $\mathbf{x}_{*}$ is not known. This
scenario can occur when a mobile user relies on a low-quality localization
system and reports an erroneous location estimate to the base station.
The wrong location has impact on the predicted posterior distribution
since the predicted mean $\mu(\mathbf{z}_{*})$ will differ from the
correct mean $\mu(\mathbf{x}_{*})$. In addition, $\mathbf{k}_{*}$
will contain erroneous entries: the $j$-th entry will be too small
when $\Vert\mathbf{z}_{*}-\mathbf{x}_{j}\Vert>\Vert\mathbf{x}_{*}-\mathbf{x}_{j}\Vert$
and too large when $\Vert\mathbf{z}_{*}-\mathbf{x}_{j}\Vert<\Vert\mathbf{x}_{*}-\mathbf{x}_{j}\Vert$.
This will affect both the posterior mean (\ref{eq:mu_pred_classicGP})
and variance (\ref{eq:var_pred_classicGP}). In the case were training
locations are also unknown, i.e., $\mathbf{Z}\neq\mathbf{X}$, and
$\mathbf{z}_{*}\neq\mathbf{x}_{*}$, these effects are further exacerbated
by the improper learning of $\boldsymbol{\theta}$.

\section{Channel Prediction with Uncertain GP \label{sec:uGP}}

In the previous section, we have argued that cGP is unable to learn
and predict properly when training or test locations are not known
exactly, especially when location error statistics are heterogeneous.
In this section, we explore several possibilities to explicitly incorporate
location uncertainty. We recall that $\mathcal{U}$ denotes the set
of all distributions over the locations in the environment $\mathcal{A}$,
while $\mathcal{X}\subset\mathcal{U}$ represents the delta Dirac
distributions over the positions and has a one-to-one mapping to $\mathcal{A}$.

We will describe three approaches. First, a Bayesian approach where
the uncertain input (i.e., the uncertain location) is marginalized,
leading to a non-Gaussian output (i.e., the received power) distribution.
Second, we derive a Gaussian approximation of the output distribution
through moment matching and detail the corresponding learning and
prediction expressions. From these expressions, the concepts of expected
mean function and expected covariance function naturally appear. Finally,
we discuss uncertain GP, which is a Gaussian process with input $\mathbf{u}$
from input set $\mathcal{U}$ and output $y$. We will relate these
three approaches in a unified view. For each approach, we detail the
quality of the solution and the computational complexity. We note
that other approaches exist, e.g., through linearizing the output
around the mean of the input \cite{mchutchon2011gaussian,girard2003learning},
but they are limited to mildly non-linear scenarios.

\subsection{Bayesian Approach}

In a Bayesian context, we learn and predict by integrating the respective
distributions over the uncertainty of the training and test locations.
As this method will involve Monte Carlo integration, we will refer
to it as Monte Carlo GP (MCGP).

\subsubsection{Learning\label{sub:MCGPLearning}}

Given the training database $\{\mathbf{U},\mathbf{y}\}$, the likelihood
function with uncertain training locations $p(\mathbf{y}\vert\mathbf{U},\boldsymbol{\theta})$
is obtained by integrating%
\footnote{For the sake of notation, all integrals in this section are written
as indefinite integrals, however they should be understood as definite
integrals over appropriate sets.%
} $p(\mathbf{y}\vert\mathbf{X},\boldsymbol{\theta})$ over the random
training locations:
\begin{equation}
p(\mathbf{y}\vert\mathbf{U},\boldsymbol{\theta})=\int p(\mathbf{y}\vert\mathbf{X},\boldsymbol{\theta})\, p(\mathbf{X})\,\mathrm{d\mathbf{X}},\label{eq:modified_likelihood}
\end{equation}
where $p(\mathbf{X})=\prod_{i=1}^{N}p(\mathbf{x}_{i})$. As there
is generally no closed-form expression for the integral (\ref{eq:modified_likelihood}),
we resort to a Monte Carlo approach by drawing $M$ i.i.d.~samples
$\mathbf{X}^{(m)}\sim p(\mathbf{X})$, $1\le m\le M$ so that
\begin{align}
p(\mathbf{y}\vert\mathbf{U},\boldsymbol{\theta}) & \approx\frac{1}{M}\sum_{m=1}^{M}p(\mathbf{y}\vert\mathbf{X}^{(m)},\boldsymbol{\theta})\nonumber \\
 & =\frac{1}{M}\sum_{m=1}^{M}\mathcal{N}(\mathbf{y};\mathbf{\mathbb{\boldsymbol{\mu}}}(\mathbf{X}^{(m)}),\mathbf{K}^{(m)}),\label{eq:MClearningSummation}
\end{align}
where $[\mathbf{K}^{(m)}]_{ij}=C(\mathbf{x}_{i}^{(m)},\mathbf{x}_{j}^{(m)})+\sigma_{n}^{2}\,\delta_{ij}$
and $\mathbf{\mathbb{\boldsymbol{\mu}}}(\mathbf{X}^{(m)})=[\mu(\mathbf{x}_{1}^{(m)}),\mu(\mathbf{x}_{2}^{(m)}),\ldots,\mu(\mathbf{x}_{N}^{(m)})]^{\mathrm{T}}$.
Finally, an estimate of $\boldsymbol{\theta}$ can be found by minimizing
the negative log-likelihood function
\begin{align}
\hat{\boldsymbol{\theta}} & =\arg\underset{\boldsymbol{\theta}}{\min}\{-\log(p(\mathbf{y}\vert\mathbf{U},\boldsymbol{\theta}))\},\label{eq:MClearning}
\end{align}
which has to be solved numerically.
\begin{rem}
\label{rem:This-optimization-involves}This optimization involves
high computational complexity and possibly numerical instability (due
to the sum of exponentials). More importantly, a good estimate of
$\boldsymbol{\theta}$ can only be found if a sample $\mathbf{X}^{(m)}$
is generated that is close to the true locations $\mathbf{X}$. Due
to the high dimensionality \cite[Section 29.2]{mackay2003information},
this is unlikely, even for large $M$. Hence, (\ref{eq:MClearning})
will lead to poor estimates of $\hat{\bm{\theta}}$.
\end{rem}

\subsubsection{Prediction}

Given the training database $\{\mathbf{U},\mathbf{y}\}$ and $\hat{\bm{\theta}}$,
we wish to determine $p(P_{\mathrm{RX}}(\mathbf{u}_{*})\vert\mathbf{U},\mathbf{y},\hat{\boldsymbol{\theta}},\mathbf{u}_{*})$
for an uncertain test location with associated distribution $p(\mathbf{x}_{*})$,
described by $\mathbf{u}_{*}$. The posterior predictive distribution
$p(P_{\mathrm{RX}}(\mathbf{u}_{*})\vert\mathbf{U},\mathbf{y},\hat{\boldsymbol{\theta}},\mathbf{u}_{*})$
is obtained by integrating $p(P_{\mathrm{RX}}(\mathbf{x}_{*})\vert\mathbf{X},\mathbf{y},\hat{\boldsymbol{\theta}},\mathbf{x}_{*})$
with respect to $\mathbf{X}$ and $\mathbf{x}_{*}$:
\begin{multline}
p(P_{\mathrm{RX}}(\mathbf{u}_{*})\vert\mathbf{U},\mathbf{y},\hat{\boldsymbol{\theta}},\mathbf{u}_{*})\\
=\int p(P_{\mathrm{RX}}(\mathbf{x}_{*})\vert\mathbf{X},\mathbf{y},\hat{\boldsymbol{\theta}},\mathbf{x}_{*})\, p(\mathbf{X})\, p(\mathbf{x}_{*})\,\mathrm{d}\mathbf{X}\,\mathrm{d}\mathbf{x}_{*}.\label{eq:preditive_dist_uncertain_train}
\end{multline}
This integral is again analytically intractable. The Laplace approximation
was utilized in \cite{jadaliha2013gaussian} to solve (\ref{eq:preditive_dist_uncertain_train}),
while here we again resort to a Monte Carlo method by drawing $M$
i.i.d.~samples $\mathbf{X}^{(m)}\sim p(\mathbf{X})$ and $\mathbf{x}_{*}^{(m)}\sim p(\mathbf{x}_{*})$,
so that
\begin{align}
 & p(P_{\mathrm{RX}}(\mathbf{u}_{*})\vert\mathbf{U},\mathbf{y},\hat{\boldsymbol{\theta}},\mathbf{u}_{*})\nonumber \\
 & \approx\frac{1}{M}\sum_{m=1}^{M}p(P_{\mathrm{RX}}(\mathbf{x}_{*}^{(m)})\vert\mathbf{X}^{(m)},\mathbf{y},\hat{\boldsymbol{\theta}},\mathbf{x}_{*}^{(m)})\nonumber \\
 & =\frac{1}{M}\sum_{m=1}^{M}\mathcal{N}(P_{\mathrm{RX}}(\mathbf{x}_{*}^{(m)});\bar{P}_{\mathrm{RX}}(\mathbf{x}_{*}^{(m)}),V_{\mathrm{RX}}(\mathbf{x}_{*}^{(m)})).\label{eq:MC_UL_UT}
\end{align}
As $M$ increases, the approximate distribution will tend to the true
distribution. We refer to (\ref{eq:MClearning}) and (\ref{eq:MC_UL_UT})
as Monte Carlo GP (MCGP). From (\ref{eq:MC_UL_UT}), we can compute
the mean ($\bar{P}_{\mathrm{RX}}^{\mathrm{MC}}(\mathbf{u}_{*})$)
and the variance ($V_{\mathrm{RX}}^{\mathrm{MC}}(\mathbf{u}_{*})$)
\cite[Eq. (14.10) and Eq. (14.11)]{koller2009probabilistic} as
\begin{align}
\bar{P}_{\mathrm{RX}}^{\mathrm{MC}}(\mathbf{u}_{*}) & =\frac{1}{M}\sum_{m=1}^{M}\bar{P}_{\mathrm{RX}}(\mathbf{x}_{*}^{(m)})\\
V_{\mathrm{RX}}^{\mathrm{MC}}(\mathbf{u}_{*}) & =\frac{1}{M}\sum_{m=1}^{M}\Bigl(\bar{P}_{\mathrm{RX}}(\mathbf{x}_{*}^{(m)})-\bar{P}_{\mathrm{RX}}^{\mathrm{MC}}(\mathbf{u}_{*})\Bigr)^{2}\nonumber \\
 & +\frac{1}{M}\sum_{m=1}^{M}V_{\mathrm{RX}}(\mathbf{x}_{*}^{(m)}).
\end{align}

\begin{rem}
Prediction is numerically straightforward, though it involves the
inversion of an $N\times N$ matrix $\mathbf{K}$ for each of the
$M$ samples $\mathbf{X}^{(m)}$. In the case training locations are
known, we can utilize cGP to obtain a good estimate of $\boldsymbol{\theta}$
and efficiently and accurately compute $\bar{P}_{\mathrm{RX}}^{\mathrm{MC}}(\mathbf{u}_{*})$
and $V_{\mathrm{RX}}^{\mathrm{MC}}(\mathbf{u}_{*})$. When both training
and test locations are known, the above procedure reverts to cGP.
\end{rem}

\subsection{Gaussian Approximation}

We have seen that while MCGP can account for location uncertainty
during prediction, it will fail to deliver adequate estimates of $\boldsymbol{\theta}$
during learning (see Remark \ref{rem:This-optimization-involves}).
To address this, we can modify $p(\mathbf{y}\vert\mathbf{U},\boldsymbol{\theta})$
from (\ref{eq:modified_likelihood}) using a Gaussian approximation
through moment matching. In addition, we can also form a Gaussian
approximation of $p(P_{\mathrm{RX}}(\mathbf{u}_{*})\vert\mathbf{U},\mathbf{y},\hat{\boldsymbol{\theta}},\mathbf{u}_{*})$
for prediction. We will term this approach Gaussian approximation
GP (GAGP). The expressions that are obtained in the learning of GAGP,
namely the expectation of mean and covariance functions will be used
later in the design of uncertain GP (described in Section \ref{sub:Uncertain-GP}).

\subsubsection{Learning}

Given the training database $\{\mathbf{U},\mathbf{y}\}$, the mean
of $p(\mathbf{y}\vert\mathbf{U},\boldsymbol{\theta})$ is given by
\begin{align}
\mathbb{E}[\mathbf{y}\vert\mathbf{U},\boldsymbol{\theta}] & =\iint\mathbf{y}\, p(\mathbf{y}\vert\mathbf{X},\boldsymbol{\theta})\, p(\mathbf{X})\,\mathrm{d}\mathbf{X}\,\mathrm{d}\mathbf{y} \nonumber \\
 & =\iint(\mathbf{y}\, p(\mathbf{y}\vert\mathbf{X},\boldsymbol{\theta})\,\mathrm{d}\mathbf{y})\, p(\mathbf{X})\,\mathrm{d}\mathbf{X}\nonumber \\
 & =\int\mathbf{\mathbb{\boldsymbol{\mu}}}(\mathbf{X})\, p(\mathbf{X})\,\mathrm{d}\mathbf{X}\nonumber \\
 & =\mathbf{\mathbb{\boldsymbol{\mu}}}(\mathbf{U}),\label{eq:exp_mean}
\end{align}
where $\mathbf{\boldsymbol{\mathbf{\mu}}}(\mathbf{U})=[\mu(\mathbf{u}_{1}),\mu(\mathbf{u}_{2}),\ldots,\mu(\mathbf{u}_{N})]^{\mathrm{T}}$
and $\mu(\mathbf{u}_{i})=\int\mu(\mathbf{x}_{i})\, p(\mathbf{x}_{i})\:\mathrm{d}\mathbf{x}_{i}$.
The covariance matrix of $p(\mathbf{y}\vert\mathbf{U},\boldsymbol{\theta})$
can be expressed as
\begin{align}
 & \textrm{Cov}[\mathbf{\mathbf{y},\mathbf{y}}\vert\mathbf{U},\boldsymbol{\theta}]\nonumber \\
 & =\int\mathbf{y}\mathbf{y}^{\mathrm{T}}\, p(\mathbf{y}\vert\mathbf{X},\boldsymbol{\theta})\, p(\mathbf{X})\,\mathrm{d\mathbf{X}}\,\mathrm{d}\mathbf{y}-\mathbf{\mathbb{\boldsymbol{\mu}}}(\mathbf{U})\mathbf{\mathbb{\boldsymbol{\mu}}}(\mathbf{U})^{\mathrm{T}}\nonumber \\
 & =\int\bigl(\mathbf{K}+\mathbf{\mathbb{\boldsymbol{\mu}}}(\mathbf{X})\mathbf{\mathbb{\boldsymbol{\mu}}}(\mathbf{X})^{\mathrm{T}}\bigr)\, p(\mathbf{X})\,\mathrm{d\mathbf{X}}-\mathbf{\mathbb{\boldsymbol{\mu}}}(\mathbf{U})\mathbf{\mathbb{\boldsymbol{\mu}}}(\mathbf{U})^{\mathrm{T}}\nonumber \\
 & =\mathbf{K}_{\mathrm{u}}+\Delta,\label{eq:exp_cov}
\end{align}
where $[\mathbf{\mathbf{K}_{\mathrm{u}}}]_{ij}=C_{\mathrm{u}}(\mathbf{u}_{i},\mathbf{u}_{j})+\sigma_{n}^{2}\,\delta_{ij}$
in which
\begin{equation}
C_{\mathrm{u}}(\mathbf{u}_{i},\mathbf{u}_{j})=\int C(\mathbf{x}_{i},\mathbf{x}_{j})\, p(\mathbf{x}_{i})\, p(\mathbf{x}_{j})\,\mathrm{d}\mathbf{x}_{i}\,\mathrm{d}\mathbf{x}_{j}\label{eq:expectedCovariance}
\end{equation}
 and $\Delta$ is a diagonal matrix with entries
\begin{equation}
[\Delta]_{ii}=\int\mu^{2}(\mathbf{x}_{i})p(\mathbf{x}_{i})\,\mathrm{d}\mathbf{x}_{i}-\mu^{2}(\mathbf{u}_{i}).\label{eq:DeltawithExpectedMean}
\end{equation}
We will refer to $\mu(\mathbf{u}_{i})$ and $C_{\mathrm{u}}(\mathbf{u}_{i},\mathbf{u}_{j})$
 as the \emph{expected mean} and \emph{expected covariance} function.
We can now express the likelihood function as $p(\mathbf{y}\vert\mathbf{U},\boldsymbol{\theta})\mathbf{\approx}\mathcal{N}(\mathbf{y};\mathbf{\mathbb{\boldsymbol{\mu}}}(\mathbf{U}),\mathbf{K}_{\mathrm{u}}+\Delta),$
so that $\boldsymbol{\theta}$ can be estimated by minimizing the
negative log-likelihood function
\begin{align}
\hat{\boldsymbol{\theta}} & =\arg\underset{\boldsymbol{\theta}}{\min}\Bigl\{-\log(\mathcal{N}(\mathbf{y};\mathbf{\mathbb{\boldsymbol{\mu}}}(\mathbf{U}),\mathbf{K}_{\mathrm{u}}+\Delta))\Bigr\}.\label{eq:likelihood_Gaussian_approx}
\end{align}

\begin{rem}
Learning in GAGP involves computation of the expected mean in (\ref{eq:exp_mean})
and (\ref{eq:DeltawithExpectedMean}), as well as the expected covariance
function in (\ref{eq:expectedCovariance}). These integrals are generally
again intractable, but there are cases where closed-form expression
exist \cite{Dallaire2011,girard2004approximate}. These will be discussed
in detail in Section \ref{sub:Uncertain-GP}. GAGP avoids the numerical
problems present in MCGP and will hence generally be able to provide
a good estimate of $\boldsymbol{\theta}$.
\end{rem}

\subsubsection{Prediction}

Given the training database $\{\mathbf{U},\mathbf{y}\}$ and $\hat{\bm{\theta}}$,
we approximate the predictive distribution $p(P_{\mathrm{RX}}(\mathbf{u}_{*})\vert\mathbf{U},\mathbf{y},\hat{\boldsymbol{\theta}},\mathbf{u}_{*})$
by a Gaussian with mean $\bar{P}_{\mathrm{RX}}^{\mathrm{GA}}(\mathbf{u}_{*})$
and variance $V_{\mathrm{RX}}^{\mathrm{GA}}(\mathbf{u}_{*})$. These
are given by
\begin{align}
 & \bar{P}_{\mathrm{RX}}^{\mathrm{GA}}(\mathbf{u}_{*})\nonumber \\
 & =\mathbb{E}[P_{\mathrm{RX}}(\mathbf{u}_{*})\vert\mathbf{U},\mathbf{y},\hat{\boldsymbol{\theta}},\mathbf{u}_{*}]\nonumber \\
 & =\int\bar{P}_{\mathrm{RX}}(\mathbf{x}_{*})\, p(\mathbf{X})\, p(\mathbf{x}_{*})\,\mathrm{d}\mathbf{X}\,\mathrm{d}\mathbf{x}_{*}\nonumber \\
 & =\mu(\mathbf{u}_{*})+\sum_{i=1}^{N}\int\beta_{i}\, C(\mathbf{x}_{*},\mathbf{x}_{i})\, p(\mathbf{X})\, p(\mathbf{x}_{*})\,\mathrm{d}\mathbf{X}\,\mathrm{d}\mathbf{x}_{*}.\label{eq:mean_GAGP}
\end{align}
Note that $\beta_{i}$ is itself a function of all \textbf{$\mathbf{X}$}'s
and $\mathbf{x}_{*}$. Similarly $V_{\mathrm{RX}}^{\mathrm{GA}}(\mathbf{u}_{*})$
is calculated as
\begin{align}
 & V_{\mathrm{RX}}^{\mathrm{GA}}(\mathbf{u}_{*})\nonumber \\
 & =\mathbb{E}[P_{\mathrm{RX}}^{2}(\mathbf{u}_{*})\vert\mathbf{U},\mathbf{y},\hat{\boldsymbol{\theta}},\mathbf{u}_{*}]-\bar{P}_{\mathrm{RX}}^{\mathrm{GA}}(\mathbf{u}_{*})^{2}\\
 & =\int\Bigl(V_{\mathrm{RX}}(\mathbf{x}_{*})+\bar{P}_{\mathrm{RX}}(\mathbf{x}_{*})^{2}\Bigr)\, p(\mathbf{X})\, p(\mathbf{x}_{*})\,\mathrm{d}\mathbf{X}\,\mathrm{d}\mathbf{x}_{*}\nonumber \\
 & -\bar{P}_{\mathrm{RX}}^{\mathrm{GA}}(\mathbf{u}_{*})^{2}.\label{eq:var_GAGP}
\end{align}
Note that both $\bar{P}_{\mathrm{RX}}(\mathbf{x}_{*})$ and $V_{\mathrm{RX}}(\mathbf{x}_{*})$
are functions of $\mathbf{X}$ (see (\ref{eq:mu_pred_classicGP})--(\ref{eq:var_pred_classicGP})).
\begin{rem}
Prediction in GAGP requires complex integrals to be solved in (\ref{eq:mean_GAGP})--(\ref{eq:var_GAGP})
for which no general closed-form expressions are known. Hence, a reasonable
approach is to use GAGP to learn $\hat{\bm{\theta}}$ and MCGP for
prediction.
\end{rem}

\begin{rem}
In case training locations are known, i.e., $\mathbf{U}\in\mathcal{X}$,
(\ref{eq:mean_GAGP}) reverts to
\begin{align}
\bar{P}_{\mathrm{RX}}^{\mathrm{GA}}(\mathbf{u}_{*}) & =\mu(\mathbf{u}_{*})+\sum_{i=1}^{N}\beta_{i}\int C(\mathbf{x}_{*},\mathbf{x}_{i})\, p(\mathbf{x}_{*})\,\mathrm{d}\mathbf{x}_{*}\label{eq:GAUSS_APPROX_CL_UT_mean}
\end{align}
and (\ref{eq:var_GAGP}) becomes
\begin{align}
 & V_{\mathrm{RX}}^{\mathrm{GA}}(\mathbf{u}_{*})\nonumber \\
 & =k_{**}-\sum_{i,j=1}^{N}[\mathbf{K}{}^{-1}]_{ij}\int C(\mathbf{x}_{*},\mathbf{x}_{i})\, C(\mathbf{x}_{*},\mathbf{x}_{j})\, p(\mathbf{x}_{*})\,\mathrm{d}\mathbf{x}_{*}\nonumber \\
 & +\int\mu(\mathbf{x}_{*})^{2}\, p(\mathbf{x}_{*})\,\mathrm{d}\mathbf{x}_{*}+2\sum_{i=1}^{N}\beta_{i}\Bigl(\int\mu(\mathbf{x}_{*})\, C(\mathbf{x}_{*},\mathbf{x}_{i})\,\nonumber \\
 & \times p(\mathbf{x}_{*})\,\mathrm{d}\mathbf{x}_{*}\Bigr)+\sum_{i,j=1}^{N}\beta_{i}\beta_{j}\int C(\mathbf{x}_{*},\mathbf{x}_{i})\, C(\mathbf{x}_{*},\mathbf{x}_{j})\, p(\mathbf{x}_{*})\,\mathrm{d}\mathbf{x}_{*}\nonumber \\
 & -\bar{P}_{\mathrm{RX}}^{\mathrm{GA}}(\mathbf{u}_{*})^{2},\label{eq:GAUSS_APPROX_CL_UT_var}
\end{align}
both of which can be computed in closed form, under some conditions,
when $\mu(\mathbf{x})$ is constant in $\mathbf{x}$ \cite[Section 3.4]{girard2004approximate}.
When both $\mathbf{U}\in\mathcal{X}$ and $\mathbf{u}_{*}\in\mathcal{X}$,
GAGP reverts to cGP.
\end{rem}

\subsection{Uncertain GP\label{sub:Uncertain-GP}}

While GAGP avoids the learning problems inherent to MCGP, prediction
is generally intractable. Hence, GAGP is not a fully coherent approach
to deal with location uncertainty. To address this, we consider a
new type of GP (uGP), which \emph{operates directly on the location
distributions, rather than on the locations}. uGP involves a mean
function $\mu_{\mathrm{uGP}}(\mathbf{u}_{i}):\mathcal{U}\to\mathbb{R}$
and a positive semidefinite covariance function $C_{\mathrm{uGP}}(\mathbf{u}_{i},\mathbf{u}_{j}):\,\mathcal{U}\times\mathcal{U}\rightarrow\mathbb{R}^{+}$,
which considers as inputs $\mathbf{u}\in\mathcal{U}$ and as outputs
$y\in\mathbb{R}$. In other words,
\begin{equation}
P_{\mathrm{RX}}(\mathbf{u}_{i})\sim\mathcal{GP}(\mu_{\mathrm{uGP}}(\mathbf{u}_{i}),C_{\mathrm{uGP}}(\mathbf{u}_{i},\mathbf{u}_{j})).
\end{equation}
The mean function is given by $\mu_{\mathrm{uGP}}(\mathbf{u}_{i})=\mathbb{E}_{\mathrm{\mathbf{x}}_{i}}[\mathbb{E}_{\Psi(\mathbf{x}_{i})}[P_{\mathrm{RX}}(\mathbf{x}_{i})]]$,
already introduced as the expected mean function in (\ref{eq:exp_mean}).
However, for the mean function to be useful in a GP context, it should
be available in closed form. As in cGP, we have significant freedom
in our choice of covariance function. Apart from all technical conditions
on the covariance function as described in \cite{rasmussen2006gaussian},
it is desirable to have a covariance function that (i) is available
in closed form; (ii) leads to decreasing correlation with increasing
input uncertainty (even when both inputs have same mean); (iii) can
account for varying amounts of input uncertainty; (iv) reverts to
a covariance function of the form (\ref{eq:generic_cov_function})
when $\mathbf{u}\in\mathcal{X}$, (v) does not depend on the mean
function $\mu(\mathbf{x})$. We will now describe the mean function
$\mu_{\mathrm{uGP}}(\mathbf{u}_{i})$ and covariance function $C_{\mathrm{uGP}}(\mathbf{u}_{i},\mathbf{u}_{j})$
in detail.

\subsubsection*{The mean function}

According to law of iterated expectations, the mean function $\mu(\mathbf{u}_{i})$
is expressed as
\begin{equation}
\mu(\mathbf{u}_{i})=L_{0}-10\,\eta\,\mathbb{E}_{\mathrm{\mathbf{x}}_{i}}[\log_{10}(\Vert\mathbf{x}_{i}\Vert)].\label{eq:mean_uncertain_input}
\end{equation}
 While there is no closed-form expression available for (\ref{eq:mean_uncertain_input}),
we can form a polynomial approximation $\sum_{j=0}^{J}a_{j}\Vert\mathbf{x}_{i}\Vert^{j}\approx\log_{10}(\Vert\mathbf{x}_{i}\Vert)$,
where the coefficients $a_{j}$ are found by least squares minimization.
For a given range of $\Vert\mathbf{x}_{i}\Vert$, this approximation
can be made arbitrarily close by increasing the order $J$. When $p(\Vert\mathbf{x}_{i}\Vert)$
is approximately Gaussian (which may be the case for $\Vert\mathbf{x}_{i}\Vert\gg0$),
$\mu(\mathbf{u}_{i})\approx L_{0}-10\,\eta\,\sum_{j=0}^{J}a_{j}\,\mathbb{E}_{\mathrm{\mathbf{x}}_{i}}[\Vert\mathbf{x}_{i}\Vert^{j}]$
can be evaluated in closed form, since all Gaussian moments are known.
See Appendix \ref{sec:Expected-Mean-Function} for details on the
approximation.

\subsubsection*{The covariance function}

While any covariance function meeting the criteria (i)--(v) listed
above can be chosen, a natural choice is (see Section \ref{sub:Classical-GP-Description})
\begin{align}
C_{\mathrm{uGP}}(\mathbf{u}_{i},\mathbf{u}_{j}) & =\mathrm{Cov}[P_{\mathrm{RX}}(\mathbf{x}_{i}),P_{\mathrm{RX}}(\mathbf{x}_{j})\vert\mathbf{u}_{i},\mathbf{u}_{j}]\nonumber \\
 & =\textrm{Cov}[y_{i},y_{j}\vert\mathbf{U},\boldsymbol{\theta}]-\delta_{ij}\sigma_{n}^{2}.
\end{align}
Unfortunately, as we can see from (\ref{eq:exp_cov}), this choice
does not satisfy criterion (v). An alternative choice is the expected
covariance function $C_{\mathrm{u}}(\mathbf{u}_{i},\mathbf{u}_{j})$
from (\ref{eq:expectedCovariance}). This choice clearly satisfies
criteria (ii), (iii), (iv), and (v). To satisfy (i), we can select
appropriate covariance functions, tailored to the distributions $p(\mathbf{x}_{i})$,
or appropriate distributions $p(\mathbf{x}_{i})$ for a given covariance
function. Examples include:
\begin{itemize}
\item Polynomial covariance functions for Gaussian $p(\mathbf{x}_{i})$
\cite{Dallaire2011,girard2004approximate}.
\item Covariance functions of the form (\ref{eq:generic_cov_function})
with $p=1$, $\mathbf{x}_{i}\in\mathbb{R}$, for Laplacian $p(\mathbf{x}_{i})$.
\item Covariance functions of the form (\ref{eq:generic_cov_function})
with $p=2$, $\mathbf{x}_{i}\in\mathbb{R}^{2}$, for Gaussian $p(\mathbf{x}_{i})$
(i.e., $p(\mathbf{x}_{i})=\mathcal{N}(\mathbf{x}_{i};\mathbf{z}_{i},\bm{\Sigma}_{i})$).
The expected covariance function is then given by \cite{Dallaire2011,girard2004approximate}
\begin{align}
 & C_{\mathrm{uGP}}(\mathbf{u}_{i},\mathbf{u}_{j})=C_{\mathrm{u}}(\mathbf{u}_{i},\mathbf{u}_{j})=\delta_{ij}\,\sigma_{\mathrm{proc}}^{2}\label{eq:uncertain_SE_Cov}\\
 & +\sigma_{\Psi}^{2}\Bigl|\mathbf{I}+d_{c}^{-2}(\bm{\Sigma}_{i}+\bm{\Sigma}_{j})(1-\delta_{ij})\Bigr|^{-1/2}\nonumber \\
 & \times\exp\left(-\frac{1}{d_{c}^{2}}(\mathbf{z}_{i}-\mathbf{z}_{j})^{\mathrm{T}}(\mathbf{I}+d_{c}^{-2}(\bm{\Sigma}_{i}+\bm{\Sigma}_{j}))^{-1}(\mathbf{z}_{i}-\mathbf{z}_{j})\right).\nonumber
\end{align}
Note that the factor $|\mathbf{I}+d_{c}^{-2}(\bm{\Sigma}_{i}+\bm{\Sigma}_{j})(1-\delta_{ij})|^{-1/2}$
ensures that inputs $i\neq j$ with the same mean (i.e., $\mathbf{z}_{i}=\mathbf{z}_{j}$)
exhibit lower correlation with increasing uncertainty. The factor
$(\mathbf{I}+d_{c}^{-2}(\bm{\Sigma}_{i}+\bm{\Sigma}_{j}))^{-1}$ ensures
that the measurements taken at locations with low uncertainty (smaller
than $d_{c}$) can be explained by a large value of $d_{c}$, while
for measurements taken at locations with high uncertainty, $C_{\mathrm{u}}(\mathbf{u}_{i},\mathbf{u}_{j})$
will be small and decreasing with increasing uncertainty.
\end{itemize}

\subsubsection{Learning}

Given the training database $\{\mathbf{U},\mathbf{y}\}$ and choosing
$\mu_{\mathrm{uGP}}(\mathbf{u}_{i})=\mu(\mathbf{u}_{i})$ and $C_{\mathrm{uGP}}(\mathbf{u}_{i},\mathbf{u}_{j})=C_{\mathrm{u}}(\mathbf{u}_{i},\mathbf{u}_{j})$,
the model parameters are found by minimizing the log-likelihood function
\begin{align}
\hat{\boldsymbol{\theta}} & =\arg\underset{\boldsymbol{\theta}}{\min}\{-\log(p(\mathbf{y}\vert\mathbf{U},\boldsymbol{\theta})\}\nonumber \\
 & =\arg\underset{\boldsymbol{\theta}}{\min}\{-\log(\mathcal{N}(\mathbf{y};\mathbf{\mathbb{\boldsymbol{\mu}}}(\mathbf{U}),\mathbf{K}_{\mathrm{u}})\}.\label{eq:loglikelihood_uGP}
\end{align}
Note that in contrast to GAGP, we have constructed uGP so that $\mathbf{\mathbb{\boldsymbol{\mu}}}(\mathbf{U})$
and $\mathbf{K}_{\mathrm{u}}$ are available in closed form, making
numerical minimization tractable.
\begin{rem}
Learning of uGP (\ref{eq:loglikelihood_uGP}) corresponds to the case
of learning (\ref{eq:likelihood_Gaussian_approx}) in GAGP for $\Delta=\mathbf{0}$
(e.g., for constant mean processes).
\end{rem}

\subsubsection{Prediction}

Let $\bar{P}_{\mathrm{RX}}(\mathbf{u}_{*})$ be the mean and $V_{\mathrm{RX}}(\mathbf{u}_{*})$
be the variance of the posterior predictive distribution $p(P_{\mathrm{RX}}(\mathbf{u}_{*})\vert\mathbf{U},\mathbf{y},\hat{\boldsymbol{\theta}},\mathbf{u}_{*})$
of uGP with uncertain training and test locations, then $p(P_{\mathrm{RX}}(\mathbf{u}_{*})\vert\mathbf{U},\mathbf{y},\hat{\boldsymbol{\theta}},\mathbf{u}_{*})=\mathcal{N}(P_{\mathrm{RX}}(\mathbf{u}_{*});\bar{P}_{\mathrm{RX}}(\mathbf{u}_{*}),V_{\mathrm{RX}}(\mathbf{u}_{*}))$.
The expressions for $\bar{P}_{\mathrm{RX}}(\mathbf{u}_{*})$ and $V_{\mathrm{RX}}(\mathbf{u}_{*})$
are now in standard GP form:
\begin{align}
\bar{P}_{\mathrm{RX}}(\mathbf{u}_{*}) & =\mu(\mathbf{u}_{*})+\mathbf{k}_{\mathrm{u}*}^{\mathrm{T}}\,\mathbf{K}_{\mathrm{u}}^{-1}\,(\mathbf{y-\boldsymbol{\mu}(\mathbf{U})})\label{eq:uncertain_mean}\\
V_{\mathrm{RX}}(\mathbf{u}_{*}) & =k_{\mathrm{u}**}-\mathbf{k}_{\mathrm{u}*}^{\mathrm{T}}\,\mathbf{K}_{\mathrm{u}}^{-1}\,\mathbf{k}_{\mathrm{u}*},\label{eq:uncertain_variance}
\end{align}
where $\mathbf{k_{\mathrm{u*}}}$ is the $N\times1$ vector of cross-covariances
$C_{\mathrm{u}}(\mathbf{u_{*}},\mathbf{u}_{i})$ between the received
power at the test distribution $\mathbf{u}_{*}$ and at the training
distribution $\mathbf{u}_{i}$, and $k_{\mathrm{u}**}$ is the a priori
variance $C_{\mathrm{u}}(\mathbf{u_{*}},\mathbf{u_{*}})$.
\begin{rem}
In case the training locations are known, i.e., $\mathbf{U}\in\mathcal{X}$,
the mean $\bar{P}_{\mathrm{RX}}(\mathbf{u}_{*})$ and the variance
$V_{\mathrm{RX}}(\mathbf{u}_{*})$ can be obtained from the expressions
(\ref{eq:uncertain_mean}) and (\ref{eq:uncertain_variance}), respectively,
by setting $\bm{\Sigma}_{i}=\mathbf{0},\forall\, i\in\{1,2,\ldots,N\}$.
Furthermore, the resulting mean $\bar{P}_{\mathrm{RX}}(\mathbf{u}_{*})$
is exactly the same as (\ref{eq:GAUSS_APPROX_CL_UT_mean}), obtained
in GAGP. However, due to a different choice of covariance function,
the predicted variance $V_{\mathrm{RX}}(\mathbf{u}_{*})$ is different
from (\ref{eq:GAUSS_APPROX_CL_UT_var}).
\end{rem}

\begin{rem}
When the test location is known, i.e., $\mathbf{u}_{*}\in\mathcal{X}$,
the mean $\bar{P}_{\mathrm{RX}}(\mathbf{x}_{*})$ and the variance
$V_{\mathrm{RX}}(\mathbf{x}_{*})$ are obtained from (\ref{eq:uncertain_mean})
and (\ref{eq:uncertain_variance}) by setting $\bm{\Sigma}_{*}=\mathbf{0}$.
\begin{figure}[t]
\psfrag{a}{\hspace{-14mm}\tiny{$ \{\mathbf{z}_{i}\!=\!\phi(\mathbf{u}_{i}),y_{i}\}_{i=1}^{N} $}}
\psfrag{b}{\tiny{$ \hat{\boldsymbol{\theta}} $}}
\psfrag{c}{\hspace{-18mm}\tiny{$ \{\mathbf{z}_{i}\!=\!\phi(\mathbf{u}_{i}),y_{i}\}_{i=1}^{N} $}}
\psfrag{d}{\hspace{0mm}\tiny{$ \mathbf{z}_{*}\!=\!\phi(\mathbf{u}_{*})$}}
\psfrag{e}{\hspace{-2mm}\tiny{$(\bar{P}_{\mathrm{RX}}(\mathbf{z}_{*}),V_{\mathrm{RX}}(\mathbf{z}_{*})) $}}
\psfrag{f}{\hspace{-2mm}\tiny{$(\bar{P}_{\mathrm{RX}}(\mathbf{u}_{*}),V_{\mathrm{RX}}(\mathbf{u}_{*})) $}}
\psfrag{g}{\hspace{-1mm}\tiny{$ \mathbf{u}_{*}$}}
\psfrag{h}{\hspace{-2mm}\hspace{-8mm}\tiny{$ \{\mathbf{u}_{i},y_{i}\}_{i=1}^{N}   $}}
\psfrag{j}{\hspace{-8mm}\tiny{$ \{\mathbf{u}_{i},y_{i}\}_{i=1}^{N}   $}}
\psfrag{i}{\tiny{$ \hat{\boldsymbol{\theta}}$}}

\psfrag{m}{\hspace{-4mm}\footnotesize{Learn}}
\psfrag{n}{\hspace{-4mm}\footnotesize{Learn}}
\psfrag{o}{\hspace{-4mm}\footnotesize{Predict}}
\psfrag{p}{\hspace{-4mm}\footnotesize{Predict}}
\psfrag{l}{\footnotesize{cGP}}
\psfrag{k}{\footnotesize{uGP}}
\psfrag{q}{\hspace{-4mm}\footnotesize{Database}}
\psfrag{r}{\hspace{-5mm}\tiny{$ \{\mathbf{u}_{i},y_i\}_{i=1}^{N} $}}

\includegraphics[width=1\columnwidth]{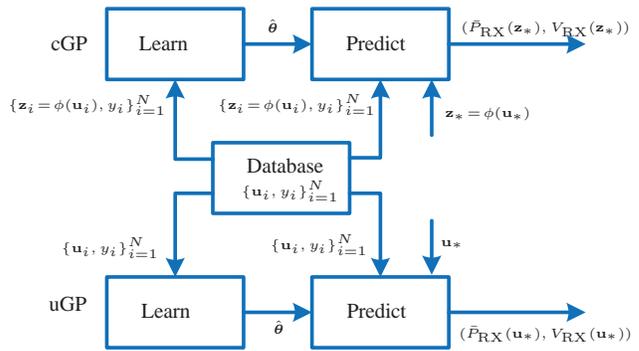}\protect\caption{\label{fig:Learning-and-testing}Learning and prediction phases of
cGP and uGP. The difference in learning in uGP compared to cGP is
that it considers location uncertainty of the nodes. The estimated
model parameters $\hat{\boldsymbol{\theta}}$ are derived during the
learning phase and are generally different in cGP compared to uGP.
The mean $\bar{P}_{\mathrm{RX}}(\mathbf{z}_{*})$ and variance $V_{\mathrm{RX}}(\mathbf{z}_{*})$
of the posterior predictive distribution in cGP corresponds to a location
$\mathbf{z}_{*}$ extracted from $\mathbf{u}_{*}$, which in turn
represents $p(\mathbf{x}_{*})$. In contrast, the mean $\bar{P}_{\mathrm{RX}}(\mathbf{u}_{*})$
and variance $V_{\mathrm{RX}}(\mathbf{u}_{*})$ of the posterior predictive
distribution in uGP pertains to the entire location distribution represented
by $\mathbf{u}_{*}$.}
\end{figure}
\begin{figure}[t]
\psfrag{U}{$\mathcal{U}$}
\psfrag{X}{$\mathcal{X}$}
\psfrag{N}{$\mathcal{N}$}
\psfrag{N}{\hspace{-21mm}\small{Gaussian output dist.}}
\psfrag{P}{$\mathcal{P}$}
\psfrag{P}{\hspace{-17mm}\small{all output dist.}}
\psfrag{text1}{\hspace{4mm}\small{cGP}}
\psfrag{text2}{$\phi$}
\psfrag{text3}{\small{MCGP}}
\psfrag{text4}{\small{GA}}
\psfrag{text5}{\small{GAGP}}
\psfrag{text6}{\small{uGP}}
\psfrag{input}{input set}
\psfrag{output}{output set}

\includegraphics[width=1\columnwidth]{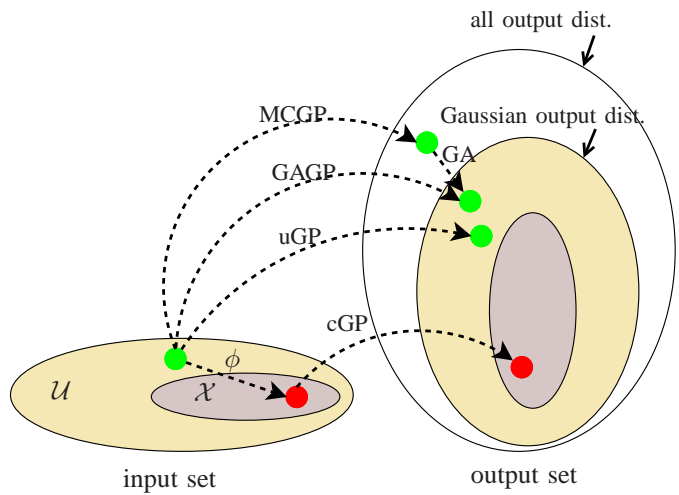}\protect\caption{\label{fig:Relation-between-cGP-uGP}Relation between cGP, MCGP, GAGP,
and uGP. All methods are equivalent when the input is limited to $\mathcal{X}$
(grey shaded area). }
\end{figure}

\end{rem}

\subsection{Unified View}

We are now ready to recap the main differences between cGP and uGP,
and to provide a unified view of the four methods (cGP, MCGP, GAGP,
and uGP). Fig.~\ref{fig:Learning-and-testing} describes the main
processes in uGP and cGP, along with the inputs and outputs during
the learning and prediction processes. The four methods are depicted
in Fig.~\ref{fig:Relation-between-cGP-uGP}: all four methods revert
to cGP when training and predictions occur in $\mathcal{X}$, i.e.,
when there is no uncertainty about the locations. MCGP is able to
consider general input distributions in $\mathcal{U}$, but leads
to non-Gaussian output distributions. Through a Gaussian approximation
of these output distributions, GAGP can consider general inputs and
directly determine a Gaussian output distribution. Both of these approaches
(MCGP and GAGP) have in common that they treat the process with input
$\mathbf{x}\in\mathcal{A}$ as a GP. In contrast, uGP treats the process
with input $\mathbf{u}\in\mathcal{U}$ as a GP. This allows for a
direct mapping from inputs in $\mathcal{U}$ to Gaussian output distributions.
In terms of tractability for learning and prediction, the four methods
are compared in Table \ref{tab:Comparision-of-benifits}. We see that
among all four methods, uGP combines tractability with good performance.

\begin{table}[tbh]
\protect\caption{\label{tab:Comparision-of-benifits}Comparison of tractability for
cGP, MCGP, GAGP, and uGP in learning and prediction. }

\centering

\begin{tabular}{|c|c|c|}
\hline
\textbf{Method} & \textbf{Learning} & \textbf{Prediction}\tabularnewline
\hline
\hline
cGP & tractable, poor quality & closed-form, poor quality\tabularnewline
\hline
MCGP & complex, poor quality & tractable\tabularnewline
\hline
GAGP & tractable in some cases & intractable\tabularnewline
\hline
uGP & tractable by design & closed-form\tabularnewline
\hline
\end{tabular}
\end{table}

\section{Numerical Results and Discussion \label{sec:Numerical-Results}}

In this section, we show learning and prediction results of cGP, uGP,
and MCGP with uncertainty in training or test locations. In Section
\ref{sub:Resource-allocation-example}, we describe a resource allocation
problem, where communication rates are predicted at future locations
using cGP and uGP, in the presence of location uncertainty during
training. The numerical analysis carried in this section is based
on simulated channel measurements according to the model outlined
in Section \ref{sec:System-Model}.

\begin{table}[tbh]
\centering\protect\caption{\label{tab:Simulation-Parameters-1}Simulation Parameters}

\begin{tabular}{|c|c|}
\hline
\textbf{Parameter} & \textbf{Value}\tabularnewline
\hline
\hline
$\eta$ & 2.5\tabularnewline
\hline
$\sigma_{n}$ & 0.01\tabularnewline
\hline
$d_{c}$ & 15 m\tabularnewline
\hline
\end{tabular}%
\begin{tabular}{|c|c|}
\hline
\textbf{Parameter} & \textbf{Value}\tabularnewline
\hline
\hline
$M$ & 300\tabularnewline
\hline
$L_{0}$ &  -10 dBm\tabularnewline
\hline
$\sigma_{\Psi}$ & 10 dB\tabularnewline
\hline
\end{tabular}
\end{table}

\subsection{Simulation Setup}

A geographical region $\mathcal{A}$ is considered and a base station
is placed at the origin. A one dimensional radio propagation field
is generated with sampling locations at a resolution of 0.25 m using
an exponential covariance function $C_{\mathrm{ref}}(\mathbf{x}_{i},\mathbf{x}_{j})=\sigma_{\Psi}^{2}\exp\Bigl(-\Vert\mathbf{x}_{i}-\mathbf{x}_{j}\Vert/d_{c}\Bigr)$,
corresponding to the Gudmundson model. Small-scale fading is assumed
to have been averaged out\footnote{In the case small-scale fading is not averaged out, the proposed framework
cannot be applied. %
}.\textcolor{blue}{{} }The simulation parameters used to obtain the numerical
results are given in Table \ref{tab:Simulation-Parameters-1}. We
assume isotropic localization errors, so that $\boldsymbol{\Sigma}_{i}=\sigma_{i}^{2}\,\mathbf{I}$.
To capture the effect of heterogeneous location errors, we draw the
location error standard deviations from an exponential distribution,
i.e., $\sigma_{i}\overset{\mathrm{i.i.d.}}{\sim}\mathrm{Exp}(\lambda)$,
where $\lambda$ is the average location error standard deviation.
For cGP and MCGP, in order to not provide any unfair advantage to
uGP, we use a covariance function of the form (\ref{eq:generic_cov_function})
with $p=1$, in order to match the true covariance function $C_{\mathrm{ref}}(\mathbf{x}_{i},\mathbf{x}_{j})$.
For uGP, we use (\ref{eq:uncertain_SE_Cov}). Since uGP exhibits a
mismatch in the covariance function, we absorb this mismatch in $\sigma_{\mathrm{proc}}$,
which is learned offline (more on this in Appendix \ref{sec:Learning}).
We assume nodes know $\sigma_{n}$ and $L_{0}$, which be inferred
using standard methods \cite{mardia1984maximum,Kitanidis1983Statistical,mostofi2010estimation},
so they are not included in the learning process.

\subsection{Learning Under Location Uncertainty}

\begin{figure*}[tbh]
\begin{centering}
\subfloat[]{\begin{centering}
\includegraphics[width=0.5\textwidth,height=0.25\textheight,keepaspectratio]{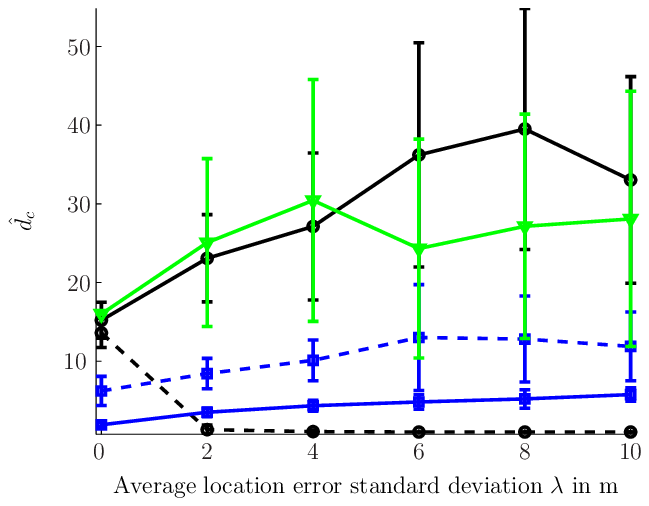}
\par\end{centering}

}\quad\subfloat[]{\begin{centering}
\includegraphics[width=0.5\textwidth,height=0.25\textheight,keepaspectratio]{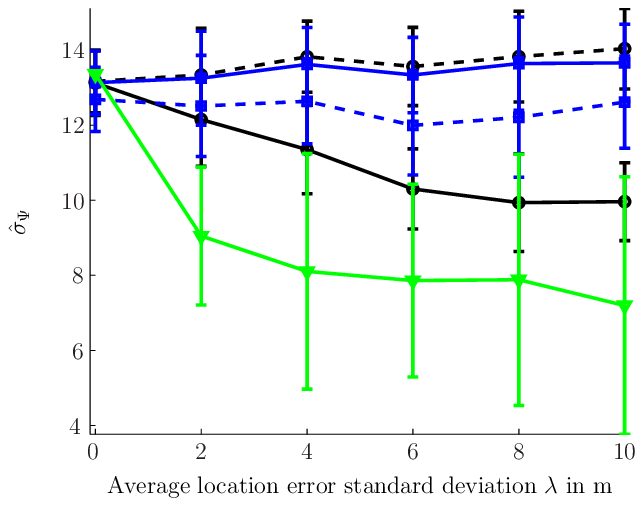}
\par\end{centering}

}
\par\end{centering}

\begin{centering}
\subfloat[]{\begin{centering}
\includegraphics[width=0.5\textwidth,height=0.25\textheight,keepaspectratio]{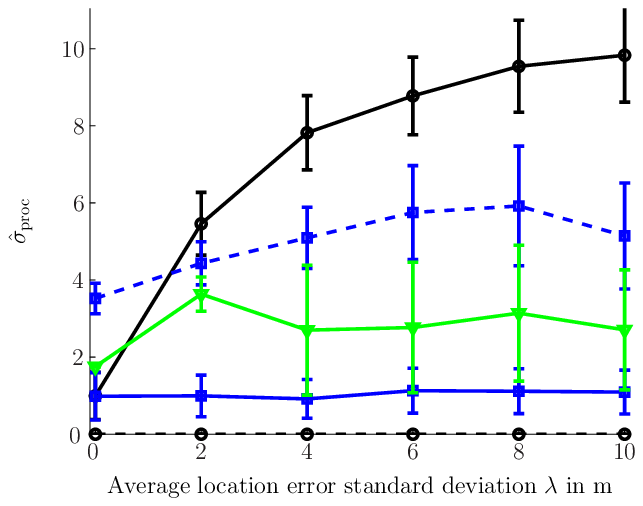}
\par\end{centering}

}\quad\subfloat[]{\begin{centering}
\includegraphics[width=0.5\textwidth,height=0.25\textheight,keepaspectratio]{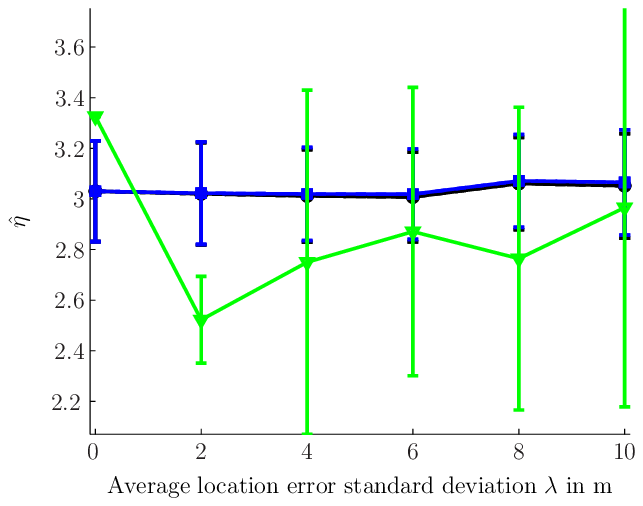}
\par\end{centering}

}
\par\end{centering}

\begin{centering}
\includegraphics[scale=3.5]{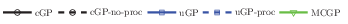}
\par\end{centering}

\protect\caption{\label{fig:Impact_location_error_learning}Impact of location uncertainty
on learning the hyperparameters using cGP, uGP, and MCGP. The hyperparameters
are estimated for each value of the mean location error standard deviation
and for 40 realizations of the channel field. Results shown are the
mean estimate of the hyperparameters and error bars with one standard
deviation. Impact of location uncertainty in shown when estimating:
(a) $d_{c}$, (b) $\sigma_{\Psi}$, (c) $\sigma_{\mathrm{proc}}$,
(d) $\eta$. }
\end{figure*}

Fig.~\ref{fig:Impact_location_error_learning} depicts the impact
of location uncertainty on the learning of hyperparameters $[d_{c},\sigma_{\Psi},\sigma_{\mathrm{proc}},\eta]$
for cGP, uGP, and MCGP. The learning of the hyperparameters is detailed
in Appendix \ref{sec:Learning}.

\subsubsection{cGP}

We first consider a variant of cGP, denoted as cGP-no-proc, in which
$\sigma_{\mathrm{proc}}$ is fixed to zero. In cGP-no-proc, when $\lambda=0$,
the estimate $\hat{d}_{c}$ is non-zero. However, it can be observed
in Fig.~\ref{fig:Impact_location_error_learning} (a), that with
increase in $\lambda$, $\hat{d}_{c}$ decreases quickly to zero.
Hence, cGP-no-proc will model the GP as a white process with high
variance $\hat{\sigma}_{\Psi}^{2}$ and thus cannot handle the location
uncertainty. On the other hand, in cGP where we estimate $\sigma_{\mathrm{proc}}$,
$\hat{\sigma}_{\mathrm{proc}}$ absorbs part of location uncertainty
(see Fig.~\ref{fig:Impact_location_error_learning} (c)). Consequently,
the part of the observations that must be explained through $\sigma_{\Psi}$
is reduced, leading to a reduction of $\hat{\sigma}_{\Psi}$ with
$\lambda$. Due to this, cGP considers the measurements constitute
a slowly varying process, therefore $\hat{d}_{c}$ increases with
$\lambda$. An interesting observation is that the error bars for
$\hat{d}_{c}$ also increase with $\lambda$. Hence, among cGP-no-proc
and cGP, only cGP can reasonably deal with location uncertainty.
\begin{figure*}[tbh]
\begin{centering}
\subfloat[]{\begin{centering}
\includegraphics[width=0.45\textwidth]{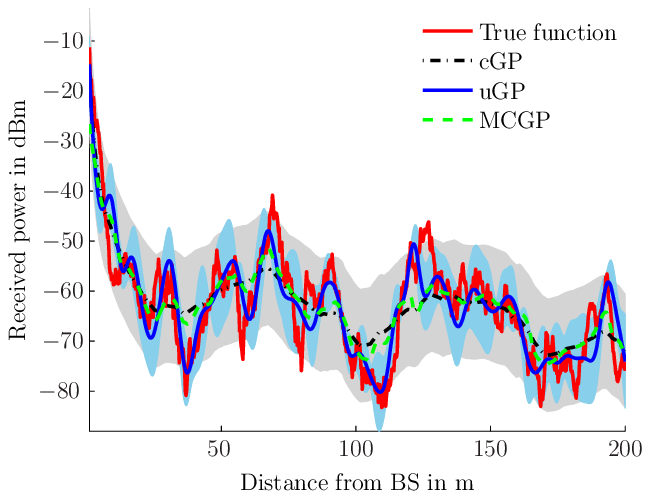}
\par\end{centering}

}\quad\subfloat[]{\begin{centering}
\includegraphics[width=0.45\textwidth]{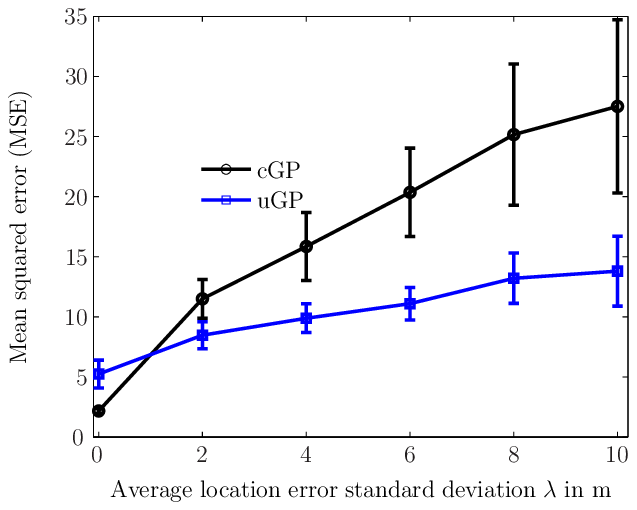}
\par\end{centering}

}
\par\end{centering}

\protect\caption{\label{fig:Perf_comp_UL_CT}Performance comparison of cGP, MCGP, and
uGP under uncertain training and certain testing locations. Inset
(a) received power prediction using uncertain training locations with
average location error of $\lambda$ = 8 m and certain test locations
for single realization of a channel field. The shaded area (grey for
cGP and blue for uGP) depicts point wise predictive mean plus and
minus the predictive standard deviation, and (b) MSE performance of
cGP and uGP as a function of average location error standard deviation
$\lambda$. The MSE is averaged for each value of $\lambda$ and for
50 realizations of the channel field is shown are the mean of the
MSE and error bars with one standard deviation. The MSE is calculated
as $\frac{1}{|\mathcal{T}|}\sum_{\mathbf{x}_{*}\in\mathcal{T}}(P_{\mathrm{RX}}(\mathbf{x}_{*})-\bar{P}_{\mathrm{RX}}(\mathbf{x}_{*}))^{2}$,
where $\mathcal{T}$ is the set of test locations and $\vert\mathcal{T}\vert$
denotes its cardinality.}
\end{figure*}

\subsubsection{MCGP}

The behavior is similar to that of cGP, i.e., an increase in $\hat{d}_{c}$,
and a decrease in $\hat{\sigma}_{\Psi}$, when increasing $\lambda$.
However, $\hat{\sigma}_{\Psi}$ decreases more quickly with $\lambda$
when compared to cGP. These effects can be attributed to two causes:
first of all, the inherent problem of drawing a finite number of samples
as detailed at the end of Section \ref{sub:MCGPLearning}; secondly,
the fluctuations in the estimated path loss exponent $\hat{\eta}$
with increasing $\lambda$ (see Fig.~\ref{fig:Impact_location_error_learning}
(d)). The error bars of the estimates in this case are even higher
than in cGP. As expected, MCGP is not suitable for learning.

\subsubsection{uGP}

As mentioned before, in uGP $\sigma_{\mathrm{proc}}$ is determined
offline. The uGP model has the capability to absorb the location uncertainty
into the covariance function. Due to this flexibility, it can handle
higher values of $\lambda$ and still maintain an almost constant
$\hat{d}_{c}$ and $\hat{\sigma}_{\Psi}$ with increase in $\lambda$.
For fair comparison with cGP, we also consider the case where $\sigma_{\mathrm{proc}}$
is estimated as part of the learning, referred to as uGP-proc. It
can be observed in Fig.~\ref{fig:Impact_location_error_learning}
(c) that $\hat{\sigma}_{\mathrm{proc}}$ increases with increase in
$\lambda$. When comparing uGP-proc to uGP, we observe a lower value
of $\hat{\sigma}_{\Psi}$ and higher values of $\hat{d}_{c}$ and
$\hat{\sigma}_{\mathrm{proc}}$ for a particular value of $\lambda$.
From this, we conclude that uGP should be preferred over uGP-proc,
as it can explain the observations with smaller $\hat{\sigma}_{\mathrm{proc}}$
and leads to simpler optimization. Finally, note that the error bars
of the uGP estimates are relatively small when compared to cGP.

\subsection{Prediction Under Location Uncertainty}

Four cases can be considered, depending on whether training or testing
inputs are in $\mathcal{X}$ or $\mathcal{U}$. We will focus on the
case where \emph{either} training or test locations are uncertain,
but not both. From these, the behavior when both training and testing
inputs are in $\mathcal{U}$ can be easily understood: only uGP can
give reasonable performance among cGP, MCGP, and uGP, as the estimates
of $\bm{\theta}$ in cGP and MCGP are of poor quality.

\subsubsection{Uncertain training locations and certain testing locations}

In this case $\mathbf{u}_{i}\in\mathcal{U}$ and $\mathbf{u}_{*}\in\mathcal{X}$.
Fig.~\ref{fig:Perf_comp_UL_CT} (a) depicts the prediction results
in terms of the predictive mean and predictive standard deviation
(shown as shaded areas) for a particular realization of the channel
field. It can be observed that uGP is able to predict the received
power comparatively better than cGP and MCGP. uGP is able to estimate
the underlying channel parameters better with the expected covariance
function, which takes in to account the location uncertainty of the
nodes. In turn, this means that uGP can track the faster variations
in the channel. cGP tries to model the true function with a slow varying
process due to very high $\hat{d}_{c}$. Furthermore, cGP has higher
uncertainty in predictions due to high $\hat{\sigma}_{\mathrm{proc}}$
(see Fig.~\ref{fig:Impact_location_error_learning} (c)). On the
other hand, MCGP has slightly better prediction performance (the standard
deviation is not shown, but is slightly smaller than for cGP) compared
to cGP due to the averaging by drawing samples from the distribution
of the uncertain training locations. Averaging the prediction error
over multiple channel realizations, Fig.~\ref{fig:Perf_comp_UL_CT}
(b) shows the mean squared error (MSE) of the received power prediction
of cGP and uGP with respect to $\lambda$ (MCGP is not shown due to
its similar performance to cGP). uGP clearly outperforms cGP (except
fo $\lambda=0$) due to its better tracking of the true channel (see
Fig.~\ref{fig:Perf_comp_UL_CT} (a)) despite uncertainty on the training
locations. The reason for higher MSE in the case of $\lambda=0$ for
uGP is due to its kernel mismatch.

\subsubsection{Certain training locations and uncertain testing locations}

\begin{figure*}[tbh]
\begin{centering}
\subfloat[]{\begin{centering}
\includegraphics[width=0.45\textwidth]{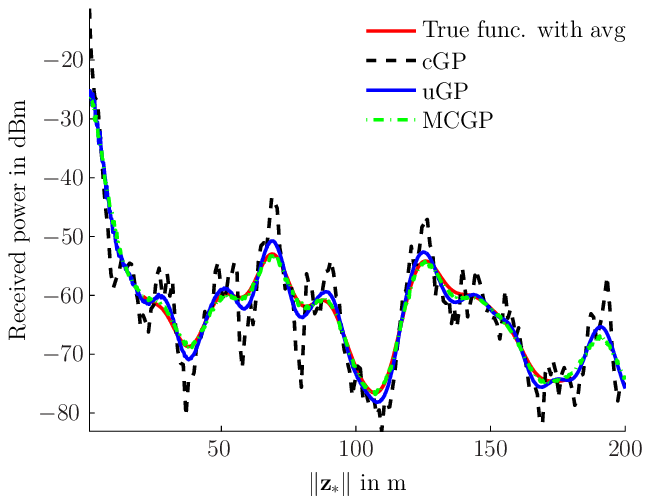}
\par\end{centering}

}\quad\subfloat[]{\begin{centering}
\includegraphics[width=0.45\textwidth]{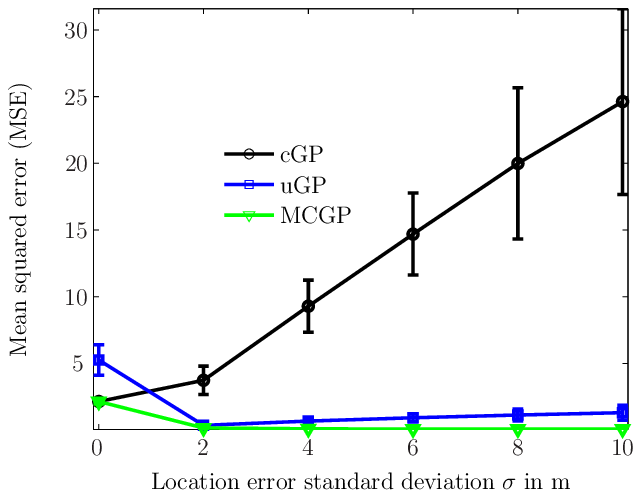}
\par\end{centering}

}
\par\end{centering}

\protect\caption{\label{fig:Perf_comp_CL_UT}Performance comparison of cGP, MCGP, and
uGP under certain training and uncertain testing locations. Inset
(a) received power prediction using certain training and uncertain
test locations with a constant location error standard deviation $\sigma=5$
m for single realization of channel field, and (b) MSE performance
of cGP, MCGP and uGP as a function of constant location error standard
deviation $\sigma$ on test locations. The MSE is averaged for each
value of $\sigma$ and for 50 realizations of the channel field is
shown are the mean of the MSE and error bars with one standard deviation.
The MSE is calculated as $\frac{1}{\vert\mathcal{T}^{u}\vert}\sum_{\mathbf{u}_{*}\in\mathcal{T}^{u}}(P_{\mathrm{RX},\mathrm{avg}}(\mathbf{u}_{*})-\bar{P}_{\mathrm{RX}}(\mathbf{u}_{*}))^{2}$,
where $\mathcal{T}^{u}$ is the set of test location distributions
and $\vert\mathcal{T}^{u}\vert$ denotes its cardinality. }
\end{figure*}

In this case $\mathbf{u}_{i}\in\mathcal{X}$ and $\mathbf{u}_{*}\in\mathcal{U}$
(with a constant location error standard deviation $\sigma$ m). Now
the performance must be assessed with respect to the expected received
power $P_{\mathrm{RX},\mathrm{avg}}(\mathbf{u}_{*})=\int P_{\mathrm{RX}}(\mathbf{x}_{*})\, p(\mathbf{x}_{*})\,\mathrm{d}\mathbf{x}_{*}$,
where $p(\mathbf{x}_{*})=\mathcal{N}(\mathbf{z}_{*},\sigma^{2}\,\mathbf{I})$,
in which $\mathbf{z}_{*}$ is the mean of distribution described by
$\mathbf{u}_{*}$. An example is shown in Fig.~\ref{fig:Perf_comp_CL_UT}
(a), depicting $P_{\mathrm{RX},\mathrm{avg}}$ as a function of $\mathbf{z}_{*}$,
as well as the predictions from cGP, MCGP, and uGP. It can be observed
that uGP and MCGP follow well $P_{\mathrm{RX},\mathrm{avg}}$. Specifically,
MCGP tracks $P_{\mathrm{RX},\mathrm{avg}}$ quite closely as it is
near-optimal in this case. In contrast, cGP follows the actual received
power at $\mathbf{z}_{*}$, rather than the averaged power. This leads
to fast variations in cGP, which are not present in uGP and MCGP.
Fig.~\ref{fig:Perf_comp_CL_UT} (b) shows the MSE of the received
power prediction of cGP, MCGP, and uGP with respect to $\sigma$ when
averaging the prediction error over multiple channel realizations.
As expected, MCGP has the lower MSE than uGP and cGP. However, uGP
performs better than cGP for all considered $\sigma$, except $\sigma=0$
(due to kernel mismatch). Furthermore, the performance of uGP is very
close to that of MCGP.

\subsection{Resource Allocation Example\label{sub:Resource-allocation-example}}

\subsubsection{Scenario}

In this section, we compare cGP and uGP for a simple proactive resource
allocation scenario. We consider a user moving through a region $\mathcal{A}$
and predict the CQM at each location. The supported rate, expressed
in bits per channel use (bpu), for a user at location $\mathbf{x}_{*}$
is defined as
\begin{equation}
r(\mathbf{x}_{*})=\log_{2}\bigl(1+\mathrm{SNR}(\mathbf{x}_{*})\bigr),
\end{equation}
where $\mathrm{SNR}(\mathbf{x}_{*})=P_{\mathrm{RX}}^{\mathrm{lin}}(\mathbf{x}_{*})/W^{\mathrm{lin}}$,
is the signal-to-noise ratio at location $\mathbf{x}_{*}$, $W^{\mathrm{lin}}$
is the receiver thermal noise and $P_{\mathrm{RX}}^{\mathrm{lin}}(\mathbf{x}_{*})$
is the received power, both measured in linear scale. The average
rate in the region $\mathcal{A}$, denoted as $\bar{r}_{\mathcal{A}}^{\mathrm{ref}}$,
is defined as
\begin{equation}
\bar{r}_{\mathcal{A}}^{\mathrm{ref}}=\frac{1}{|\mathcal{A}|}\int_{\mathcal{A}}r(\mathbf{x}_{*})\mathrm{d}\mathbf{x}_{*},
\end{equation}
where $|\mathcal{A}|$ denotes area of the region $\mathcal{A}$.
The predicted rate for a user at a future location $\mathbf{x}_{*}$,
based on the predicted CQM values $(\bar{P}_{\mathrm{RX}}(\mathbf{x}_{*}),V_{\mathrm{RX}}(\mathbf{x}_{*}))$,
is defined as
\begin{equation}
r(\mathbf{x}_{*},\alpha)=\log_{2}\bigl(1+\mathrm{SNR}(\mathbf{x}_{*},\alpha)\bigr),
\end{equation}
where $\alpha\geq0$ is a confidence parameter, $\mathrm{SNR}(\mathbf{x}_{*},\alpha)=P_{\mathrm{RX}}^{\mathrm{lin}}(\mathbf{x}_{*},\alpha)/W^{\mathrm{lin}}$
and $P_{\mathrm{RX}}(\mathbf{x}_{*},\alpha)=10\,\log_{10}\bigl(P_{\mathrm{RX}}^{\mathrm{lin}}(\mathbf{x}_{*},\alpha)\bigr)=\bar{P}_{\mathrm{RX}}(\mathbf{x}_{*})-\alpha\,\bigl(V_{\mathrm{RX}}(\mathbf{x}_{*})\bigr)^{\frac{1}{2}}$
.

\subsubsection{Performance measure}

The user moves through the environment according to a known trajectory.
The base station allocates bits to each future location, proportional
to $r(\mathbf{x}_{*},\alpha)$. When the user is at location $\mathbf{x}_{*}$,
only a fraction of the bits, proportional to $\min(r(\mathbf{x}_{*},\alpha),r(\mathbf{x}_{*}))$
would be delivered. Therefore, the effective rate $r^{\mathrm{eff}}(\mathbf{x}_{*},\alpha)$
for the user at location $\mathbf{x}_{*}$ is
\begin{figure*}[tbh]
\begin{centering}
\subfloat[]{\begin{centering}
\includegraphics[width=0.45\textwidth]{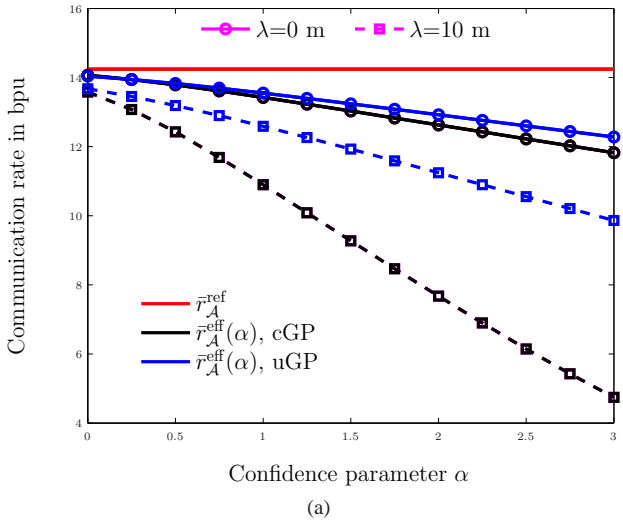}
\par\end{centering}

}\quad\subfloat[]{\begin{centering}
\includegraphics[width=0.45\textwidth]{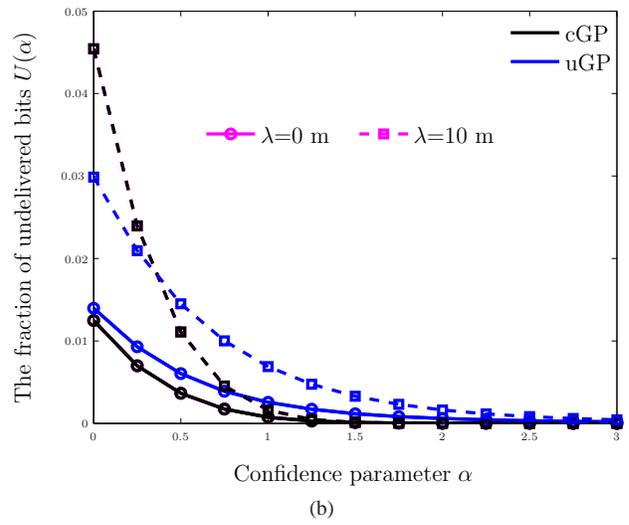}
\par\end{centering}

}
\par\end{centering}

\protect\caption{\label{fig:Resource-allocation-example}Resource allocation example
for cGP, and uGP with two different values of localization error standard
deviations ($\lambda\in\{0,10\}$ m) and for different values of the
confidence parameter $\alpha$. The results are averaged for each
value of $\lambda$ with 50 channel realizations. Inset (a) the effective
rate $\bar{r}_{\mathcal{A}}^{\mathrm{eff}}(\alpha)$, and (b) the
fraction of undelivered bits $U(\alpha)$.}
\end{figure*}
\begin{align}
r^{\mathrm{eff}}(\mathbf{x}_{*},\alpha) & =\min(r(\mathbf{x}_{*},\alpha),r(\mathbf{x}_{*})).
\end{align}
The average effective rate $\bar{r}_{\mathcal{A}}^{\mathrm{eff}}(\alpha)$
for a given confidence level $\alpha$ is then computed by spatial
average of $r^{\mathrm{eff}}(\mathbf{x}_{*},\alpha)$ over region
$\mathcal{A}$ as
\begin{align}
\bar{r}_{\mathcal{A}}^{\mathrm{eff}}(\alpha) & =\frac{1}{|\mathcal{A}|}\int_{\mathcal{A}}\, r^{\mathrm{eff}}(\mathbf{x}_{*},\alpha)\,\mathrm{d}\mathbf{x}_{*}\in[0,\bar{r}_{\mathcal{A}}^{\mathrm{ref}}].
\end{align}
When $r(\mathbf{x}_{*},\alpha)>r(\mathbf{x}_{*})$, a part of the
allocated bits cannot be delivered. The total fraction of undelivered
bits over the environment is given by
\begin{equation}
U(\alpha)=\frac{\int_{\mathcal{A}}\,(r(\mathbf{x}_{*},\alpha)-r^{\mathrm{eff}}(\mathbf{x}_{*},\alpha))\,\mathrm{d}\mathbf{x}_{*}}{\int_{\mathcal{A}}\, r(\mathbf{x}_{*},\alpha)\,\mathrm{d}\mathbf{x}_{*}}\in[0,1).
\end{equation}
Hence, $\bar{r}_{\mathcal{A}}^{\mathrm{eff}}(\alpha)$ describes the
rate that the user will receive (penalizing under-estimation of the
rate), while $U(\alpha)$ describes the loss due to lost bits (penalizing
over-estimating of the rate).

\subsubsection{Predicted communication rates with uncertain training locations}

We predict the CQM at known test locations $\mathbf{x}_{*}\in\mathcal{X}$,
based on training with uncertain locations (considering $\lambda\in\{0,10\}$
m), all within a one-dimensional region $\mathcal{A}$. The average
effective rate $\bar{r}_{\mathcal{A}}^{\mathrm{eff}}(\alpha)$ and
the fraction of undelivered bits $U(\alpha)$, as a function of $\alpha$,
are shown in Fig~\ref{fig:Resource-allocation-example} (a)--(b),
respectively. As expected, increasing $\alpha$ leads to a more conservative
allocation, thus reducing both $\bar{r}_{\mathcal{A}}^{\mathrm{eff}}(\alpha)$
and $U(\alpha)$. For a specific value of $\alpha$, increase in $\lambda$
decreases $\bar{r}_{\mathcal{A}}^{\mathrm{eff}}(\alpha)$. This is
due to the fact that with increase in $\lambda$, the mean $\bar{P}_{\mathrm{RX}}(\mathbf{x}_{*})$
is of poor quality and the variance $V_{\mathrm{RX}}(\mathbf{x}_{*})$
is high for CQM predictions.

It is evident that when $\lambda=0$, uGP and cGP attain similar performance,
both in terms of $\bar{r}_{\mathcal{A}}^{\mathrm{eff}}(\alpha)$ and
$U(\alpha)$. When $\lambda$ is increased to 10 m, cGP suffers from
a significant reduction in effective rate $\bar{r}_{\mathcal{A}}^{\mathrm{eff}}(\alpha)$,
while at the same time dropping up to 4.5 \% of the bits. This is
due to cGP's poor predictions, which are either too low (leading to
a reduction in $\bar{r}_{\mathcal{A}}^{\mathrm{eff}}(\alpha)$) or
too high (leading to an increase in $U(\alpha)$). In contrast, uGP,
which is able to track the channel well despite uncertain training,
achieves a higher effective rate, especially for high confidence values
(e.g., around 2 times higher rate for $\alpha=3$, for $U(\alpha)$
less than 0.1\%).

\section{Conclusion \label{sec:Conclusion}}

Channel quality metrics can be predicted using spatial regression
tools such as Gaussian processes (GP). We have studied the impact
of location uncertainties on GP and have demonstrated that, when heterogeneous
location uncertainties are present, the classical GP framework is
unable to (i) learn the underlying channel parameters properly; (ii)
predict the expected channel quality metric. By introducing a GP that
operates directly on the location distribution, we find uncertain
GP (uGP), which is able to both learn and predict in the presence
of location uncertainties. This translates in better performance when
using uGP for predictive resource allocation.

Possible avenues of future research include validation using real
measurements, modeling correlation of shadowing in the temporal dimension,
study of better approximations for learning with uncertain locations,
and the extension to ad-hoc networks.

\appendices{}

\section{Approximation of Expected Mean Function\label{sec:Expected-Mean-Function}}

Let $d_{i}=\Vert\mathbf{x}_{i}\Vert$ and recall from random variable
transformation theory that
\begin{equation}
\int\log_{10}(\Vert\mathbf{x}_{i}\Vert)\, p(\mathbf{x}_{i})\,\mathrm{d}\mathbf{x}_{i}=\int\log_{10}(d_{i})\, p(d_{i})\,\mathrm{d}d_{i}.\label{eq:log_integral}
\end{equation}
We assume $p(\mathbf{x}_{i})=\mathcal{N}(\mathbf{z}_{i},\sigma_{i}^{2}\,\mathbf{I})$,
so $p(d_{i})$ follows a Rician distribution
\begin{equation}
p(d_{i})=\frac{d_{i}}{\sigma_{i}^{2}}\,\exp\Bigl(-\frac{\Vert\mathbf{z}_{i}\Vert{}^{2}+d_{i}^{2}}{2\,\sigma_{i}^{2}}\Bigr)\, I_{0}\Bigl(\frac{\Vert\mathbf{z}_{i}\Vert\, d_{i}}{\sigma_{i}^{2}}\Bigr)\: d_{i}>0,
\end{equation}
where $I_{0}(.)$ is a modified Bessel function of zero-th order.
For $\Vert\mathbf{z}_{i}\Vert/\sigma_{i}\geq3$, $p(d_{i})$ can be
approximated as a Gaussian distribution
\begin{equation}
p_{\mathrm{Gauss}}(d_{i})=\frac{1}{\sqrt{2\,\pi\sigma_{i}^{2}}}\,\exp\Bigl(-\frac{(\Vert\mathbf{z}_{i}\Vert-d_{i})^{2}}{2\,\sigma_{i}^{2}}\Bigr).
\end{equation}
The integral (\ref{eq:log_integral}) still does not have a closed
form expression with $p_{\mathrm{Gauss}}(d_{i})$. Now approximating
the $\log_{10}(.)$ function with a polynomial function of the form
$w(d_{i})=\sum_{j=0}^{J}a_{j}\, d_{i}^{j}$ then (\ref{eq:log_integral})
can be written as
\begin{equation}
\int\log_{10}(\Vert\mathbf{x}_{i}\Vert)\, p(\mathbf{x}_{i})\,\mathrm{d}\mathbf{x}_{i}\approx\int_{-\infty}^{+\infty}w(d_{i})\, p_{\mathrm{Gauss}}(d_{i})\mathrm{\, d}d_{i},
\end{equation}
which can be computed exactly.

\section{Learning Procedure\label{sec:Learning}}

In this appendix, we detail the learning of $\boldsymbol{\theta}=[\sigma_{n},\sigma_{\mathrm{proc}},d_{c},L_{0},\eta,\sigma_{\Psi}]$
for cGP, uGP, and MCGP. We consider nodes know $\sigma_{n}$ and $L_{0}$,
therefore they are not estimated as part of the learning process.
Let the remaining set of hyperparameters be $\boldsymbol{\theta}=[\sigma_{\mathrm{proc}},d_{c},\sigma_{\Psi}]$
and $\eta$ .

\subsection*{cGP}

Based on Section \ref{sec:System-Model}, we can write the received
measurements $\mathbf{y}$ with their corresponding training locations
$\mathbf{X}$ in matrix form as
\begin{equation}
\mathbf{y}=\mathbf{1}^{\mathrm{T}}L_{0}+\mathbf{h}_{\mathrm{c}}\,\eta+\boldsymbol{\Psi}+\mathbf{n},
\end{equation}
where $\boldsymbol{\Psi}=[\Psi(\mathbf{x}_{1}),\ldots\Psi(\mathbf{x}_{N})]^{\mathrm{T}}$,
$\mathbf{n}=[n_{1},\ldots,n_{N}]^{\mathrm{T}},$ and $\mathbf{h}_{\mathrm{c}}=-10\,[\,\log_{10}(\Vert\mathbf{x}_{1}\Vert),\ldots,\log_{10}(\Vert\mathbf{x}_{N}\Vert)]^{\mathrm{T}}$.
Assuming the measurements are uncorrelated, then the least squares
estimate of the path-loss exponent can be computed as
\begin{equation}
\hat{\eta}=\bigl(\mathbf{h}_{\mathrm{c}}^{\mathrm{T}}\,\mathbf{h}_{\mathrm{c}}\bigr)^{-1}\,\mathbf{h}_{\mathrm{c}}^{\mathrm{T}}\,\bigl(\mathbf{y}-\mathbf{1}^{\mathrm{T}}L_{0}\bigr).
\end{equation}
Once the path-loss exponent is estimated, the mean component of the
received measurements can be subtracted as, $\boldsymbol{\Upsilon}_{\mathrm{c}}=\mathbf{y}-\mathbf{1}^{\mathrm{T}}L_{0}-\mathbf{h}_{\mathrm{c}}\,\hat{\eta}$.
Then, $\boldsymbol{\Upsilon}_{\mathrm{c}}$ becomes a zero-mean Gaussian
process. Now the likelihood function (\ref{eq:GPtraining}) becomes
$l(\boldsymbol{\theta})=p(\mathbf{\boldsymbol{\Upsilon}}_{\mathrm{c}}\vert\mathbf{X},\boldsymbol{\theta})=\mathcal{N}(\mathbf{\boldsymbol{\Upsilon}}_{\mathrm{c}};\mathbf{0},\mathbf{K})$.
The hyperparameters $\boldsymbol{\theta}$ are estimated by minimizing
negative logarithm of $l(\boldsymbol{\theta})$
\begin{align}
\hat{\boldsymbol{\theta}} & =\arg\underset{\boldsymbol{\theta}}{\min}\{-\log(p(\boldsymbol{\Upsilon}_{\mathrm{c}}\vert\mathbf{X},\boldsymbol{\theta})\}\nonumber \\
 & =\arg\underset{\boldsymbol{\theta}}{\min}\Bigl\{\log|\mathbf{K}|+\boldsymbol{\Upsilon}_{\mathrm{c}}^{\mathrm{T}}\,\mathbf{K}{}^{-1}\,\boldsymbol{\Upsilon}_{\mathrm{c}}\Bigr\}.\label{eq:new_likelihood}
\end{align}
We calculate the variance of the process $\boldsymbol{\Upsilon}_{\mathrm{c}}$
as $\sigma_{\mathrm{Tot}}^{2}=1/N\sum_{i=1}^{N}[\boldsymbol{\Upsilon}_{\mathrm{c}}]_{i}^{2}$.
The variance of the process should be captured by the hyperparameters
$\sigma_{\mathrm{proc}}$, $\sigma_{n}$, and $\sigma_{\Psi}$. We
define $\sigma_{\mathrm{proc}}^{2}=\sigma_{\mathrm{Tot}}^{2}-\sigma_{n}^{2}-\sigma_{\Psi}^{2}$,
as a result $l(\boldsymbol{\theta})$ becomes a function of only $d_{c}$
and $\sigma_{\Psi}$. We solve (\ref{eq:new_likelihood}) and find
$\hat{d}_{c}$ and $\hat{\sigma}_{\Psi}$ by an exhaustive grid search.
Once $\hat{d}_{c}$ and $\hat{\sigma}_{\Psi}$ are found, then $\hat{\sigma}_{\mathrm{proc}}$
can be calculated as $\hat{\sigma}_{\mathrm{proc}}^{2}=\hat{\sigma}_{\mathrm{Tot}}^{2}-\sigma_{n}^{2}-\hat{\sigma}_{\Psi}^{2}$.

\subsection*{uGP}

In this case, the path-loss exponent is estimated as
\begin{equation}
\hat{\eta}=\bigl(\mathbf{h}_{\mathrm{u}}^{\mathrm{T}}\,\mathbf{h}_{\mathrm{u}}\bigr)^{-1}\,\mathbf{h}_{\mathrm{u}}^{\mathrm{T}}\,\bigl(\mathbf{y}-\mathbf{1}^{\mathrm{T}}L_{0}\bigr),
\end{equation}
where $\mathbf{h}_{\mathrm{u}}=-10\,[\mathbb{E}_{\mathrm{\mathbf{x}}_{1}}[\log_{10}(\Vert\mathbf{x}_{1}\Vert),\ldots,\mathbb{E}_{\mathrm{\mathbf{x}}_{N}}[\log_{10}(\Vert\mathbf{x}_{N}\Vert)]^{\mathrm{T}}$.
Once again removing the mean from the measurements, we obtain $\boldsymbol{\Upsilon}_{\mathrm{u}}=\mathbf{y}-\mathbf{1}^{\mathrm{T}}L_{0}-\mathbf{h}_{\mathrm{u}}\,\hat{\eta}.$
The hyperparameters $\boldsymbol{\theta}$ are estimated by minimizing
the modified negative log-likelihood function
\begin{align}
\hat{\boldsymbol{\theta}} & =\arg\underset{\boldsymbol{\theta}}{\min}\{-\log(p(\boldsymbol{\Upsilon}_{\mathrm{u}}\vert\mathbf{U},\boldsymbol{\theta})\}\nonumber \\
 & =\arg\underset{\boldsymbol{\theta}}{\min}\Bigl\{\log|\mathbf{K}_{\mathrm{u}}|+\boldsymbol{\Upsilon}_{\mathrm{u}}^{\mathrm{T}}\,\mathbf{K}_{\mathrm{u}}{}^{-1}\,\boldsymbol{\Upsilon}_{\mathrm{u}}\Bigr\}.\label{eq:new_likelihood-uGP}
\end{align}
Again, $\sigma_{\mathrm{Tot}}^{2}=1/N\sum_{i=1}^{N}[\boldsymbol{\Upsilon}_{\mathrm{u}}]_{i}^{2}$,
is the variance of the process. As a result, $\hat{\sigma}_{\Psi}$
becomes $\hat{\sigma}_{\Psi}^{2}=\sigma_{\mathrm{Tot}}^{2}-\sigma_{n}^{2}-\hat{\sigma}_{\mathrm{proc}}^{2}$
and due to this $l(\boldsymbol{\theta})$ is now only a function of
$d_{c}$. We solve (\ref{eq:new_likelihood-uGP}) and find $\hat{d}_{c}$
by an exhaustive grid search.

The learning process can be simplified for uGP: since $\sigma_{\mathrm{proc}}$
only captures kernel mismatch irrespective of the location uncertainty
and path loss, the value of $\hat{\sigma}_{\mathrm{proc}}$ can be
obtained off-line with noise-free training locations by performing
learning as in the case of cGP, but with a covariance function of
the form (\ref{eq:generic_cov_function}) for $p=2$. This approach
gives an advantage to cGP and thus makes the comparison between uGP
and cGP more fair for all values of $\lambda\ge0$.

\subsection*{MCGP}

It is no longer feasible to estimate $\eta$ first and subtract to
make the process zero mean, because of summation in the Monte Carlo
integration (\ref{eq:MClearningSummation}). Therefore, we optimize
(\ref{eq:MClearning}) with respect to the hyperparameters $\eta$
and $\boldsymbol{\theta}$ using $\mathtt{fminsearch}$ function of
Matlab.

\section*{Acknowledgment}

The authors would like to thank Ido Nevat, Lennart Svensson, Ilaria
Malanchini, and Vinay Suryaprakash for their feedback on the manuscript.

\bibliographystyle{IEEEtran}

\end{document}